\documentclass[aip,jcp,amsmath,amssymb,floatfix,reprint,doi=true,citeautoscript]{revtex4-2}

\usepackage[british]{babel}
\usepackage[utf8]{inputenc}
\usepackage{graphicx}
\usepackage{placeins}
\usepackage{wrapfig}
\usepackage{bibunits}
\usepackage{xr-hyper}
\usepackage[colorlinks,allcolors=black,citecolor=blue,urlcolor=blue]{hyperref}
\usepackage{tipa}
\usepackage{siunitx}
\usepackage{dcolumn}
\usepackage{gensymb}

\hypersetup{colorlinks=true, urlcolor=blue}

\newcommand{\vect}[1]{\mathbf{#1}}

\newcommand*{\revPBET}{revPBE0\nobreakdashes-D3/\allowbreak TZV2P/\allowbreak GTH}
\newcommand*{\revPBEQGTH}{revPBE0\nobreakdashes-D3/\allowbreak def2\nobreakdashes-QZVP/\allowbreak GTH}
\newcommand*{\revPBEQ}{revPBE0\nobreakdashes-D3/\allowbreak def2\nobreakdashes-QZVP/\allowbreak AE}
\newcommand*{\revPBElone}{revPBE0\nobreakdashes-D3}
\newcommand*{\Bninety}{\textomega B97X\nobreakdashes-rV/\allowbreak def2\nobreakdashes-QZVP/\allowbreak AE}
\newcommand*{\Bninetylone}{\textomega B97X\nobreakdashes-rV}
\newcommand*{\BninetyVlone}{\textomega B97X\nobreakdashes-V}
\newcommand*{\MPtwo}{MP2/\allowbreak cc\nobreakdashes-TZ/\allowbreak GTH}
\newcommand*{\ih}{$\mathrm{I_h}$}

\defaultbibliography{paper,revtex-control}
\defaultbibliographystyle{aipnum4-2}

\begin{document}


\def\mytitle{More converged, less accurate? Reassessing standard choices for ab initio water using machine learning potentials}
\title{\mytitle}

\author{Hubert Beck}
\affiliation{
Charles University, Faculty of Mathematics and Physics, Ke Karlovu 3, 121 16 Prague 2, Czech Republic
}

\author{Ondrej Marsalek}
\email{ondrej.marsalek@matfyz.cuni.cz}
\affiliation{
Charles University, Faculty of Mathematics and Physics, Ke Karlovu 3, 121 16 Prague 2, Czech Republic
}

\date{\today}

\begin{abstract}
Accurately simulating the properties of liquid water remains a central challenge in molecular simulations.
In this work, we use machine learning potentials to investigate how the convergence settings of electronic structure calculations impact the predicted structural and dynamical properties of simulated water and ice.
We evaluate the true performance of several reference methods in classical and path-integral molecular dynamics.
When we compare a popular, computationally pragmatic revPBE0-D3 setup against a highly converged one,
our results reveal that its widely reported experimental agreement degrades.
Applying the same highly converged settings to the \textomega{}B97X-rV functional, we find an improved agreement with experimental results.
MP2 with a triple-$\zeta$ basis set commonly used for liquid water shows poor performance, which is indicative of insufficient convergence.
These findings underscore the need for fully converged reference calculations when evaluating the fundamental accuracy of electronic structure methods and developing reliable models for aqueous systems.
\end{abstract}

{\maketitle}

\begin{bibunit}

\nocite{revtex-control}

\section{Introduction}

Arguably, there is no other material that has been studied more intensely than water~\cite{Cisernos2016/10.1021/acs.chemrev.5b00644, Ceriotti2016/10.1021/acs.chemrev.5b00674, Gillan2016/10.1063/1.4944633}.
Despite its simple molecular structure, it has been in the center of attention for many decades and to this date it continues to draw interest and spark debates~\cite{Clark2010/10.1080/00268971003762134, Riera2019/10.1039/C9SC03291F, Gallo2016/10.1021/acs.chemrev.5b00750}.
Its role as a solvent for many processes, including biological ones~\cite{Bellissent-Funel2016/10.1021/acs.chemrev.5b00664}, underscores the importance of finding efficient and reliable ways to simulate the properties of water as accurately as possible.
Furthermore, it displays interesting microscopic properties, such as a complex network of hydrogen bonds~\cite{Head-Gordon2006/10.1073/pnas.0510593103, Liu2017/10.1039/C7CP00667E, Liu2018/10.1039/C7SC04205A} or intricate dynamics~\cite{Wilkins2017/10.1021/acs.jpclett.7b00979}, as well as macroscopic properties, such as the unique behavior of its density~\cite{Morawietz2016/10.1073/pnas.1602375113, deHijes2024/10.1063/5.0227514}.
Accurately modeling experimental properties such as the radial distribution function (RDF), the density, or the diffusion coefficient across the phase diagram is a field of active research~\cite{Willow2016/10.1021/acs.jpclett.5b02430, Pestana2017/10.1039/C6SC04711D, Yao2018/10.1021/acs.jctc.7b00846, Cheng2019/10.1073/pnas.1815117116, Li2025/10.4208/cicc.2025.88.01}.

Density functional theory (DFT) has been the main tool in the quest for accurate liquid water properties from first principles and different levels of theory have been extensively tested~\cite{Ceriotti2013/10.1073/pnas.1308560110, Galib2017/10.1063/1.4986284, Riera2019/10.1039/C9SC03291F, Villard2024/10.1039/D3SC05828J, Pestana2018/10.1021/acs.jpclett.8b02400}.
In the last decade, machine learning potentials (MLPs) have increasingly gained popularity, driven by their ability to accurately reproduce the potential energy surface (PES) of high-level ab initio electronic structure methods at a substantially lower computational cost~\cite{Behler2007/10.1103/PhysRevLett.98.146401, MartinBarrios2024/10.1002/qua.27389}.
As such, they are perfect for driving long molecular dynamics (MD) simulations and, therefore, have been trained to many different electronic structure methods to find the ideal setup for water~\cite{Morawietz2016/10.1073/pnas.1602375113, Cheng2019/10.1073/pnas.1815117116, Liu2022/10.1021/acs.jpca.2c00601, Chen2023/10.1021/acs.jctc.2c01203, deHijes2024/10.1063/5.0227514, Li2025/10.4208/cicc.2025.88.01, Dao2026/10.26434/chemrxiv.15000644/v1}.
These simulations have been used to study the properties of water in its many different phases as well as its interactions with solutes or interfaces~\cite{Schran2021/10.1073/PNAS.2110077118, Omranpour2024/10.1063/5.0201241, Advincula2026/10.1039/D5FD00165J}.

MLPs enabled the study of the importance of including dispersion interactions in water simulations much more thoroughly than with ab initio methods alone~\cite{Schmidt2009/10.1021/jp901990u, Morawietz2016/10.1073/pnas.1602375113}.
While long-range dispersion interactions are not directly included in most DFT exchange--correlation functionals, they can be added approximately through dispersion correction methods~\cite{Grimme2011/10.1002/wcms.30, Klimes2012/10.1063/1.4754130}.
Some functionals include dispersion interactions directly through a non-local correlation term in the functional itself~\cite{Dion2004/10.1103/PhysRevLett.92.246401, Vydrov2010/10.1063/1.3521275}.
Alternatively, correlated methods, such as second order Møller--Plesset perturbation theory (MP2), random phase approximation (RPA), or coupled cluster methods, naturally capture long-range electron correlations, including dispersion interactions.
Running MD simulations at this level of theory~\cite{Liu2022/10.1021/acs.jpca.2c00601, Chen2023/10.1021/acs.jctc.2c01203, Yu2023/10.1021/acs.jpclett.3c01791, Li2025/10.4208/cicc.2025.88.01, ONeill2025/10.1021/acs.jctc.5c01377} was previously only attainable at enormous computational cost~\cite{DelBen2013/10.1021/jz401931f, DelBen2015/10.1063/1.4927325}, further highlighting the advantages of MLPs.

With and without the use of MLPs, nuclear quantum effects (NQEs) can be included in MD simulations through imaginary-time path integrals~\cite{Markland2018/10.1038/s41570-017-0109}.
NQEs influence many aspects of aqueous systems, such as structure, vibrational properties, or diffusion~\cite{Paesani2006/10.1063/1.2386157, Paesani2009/10.1021/jp810590c, Ceriotti2013/10.1073/pnas.1308560110, Wang2014/10.1063/1.4894287, Ceriotti2016/10.1021/acs.chemrev.5b00674, Marsalek2017/10.1021/acs.jpclett.7b00391, Li2025/10.4208/cicc.2025.88.01}.
Competing NQEs on intermolecular and intramolecular interactions make the resulting properties particularly sensitive to the details of the underlying PES~\cite{Habershon2009/10.1063/1.3167790, Li2011/10.1073/pnas.1016653108, Wilkins2017/10.1021/acs.jpclett.7b00979}.
This can result in even seemingly similar models yielding a broad range of outcomes.

One of the most popular exchange--correlation functionals for water simulations is the hybrid functional \mbox{revPBE0}, which is a revPBE generalized gradient approximation (GGA) functional where 25\% of the exchange energy was replaced by the exact Hartree--Fock (HF) exchange energy.
When evaluated in the TZV2P basis set using pseudopotentials and combined with Grimme's D3 correction, it reproduces many important experimental findings of water, such as structural properties, the diffusion coefficient, and vibrational spectra, very well~\cite{Marsalek2017/10.1021/acs.jpclett.7b00391}.
Therefore, this setup has seen broad adaptation for simulations of liquid water~\cite{Cheng2019/10.1073/pnas.1815117116, Reinhardt2021/10.1038/s41467-020-20821-w, Sharma2021/10.1021/acs.jctc.1c00582, Kapil2022/10.1038/s41586-022-05036-x, Shen2025/10.1021/acs.jpcb.5c03479, Cunningham2026/10.1021/acs.jcim.6c00491}.
Another promising candidate for simulations of aqueous systems is \BninetyVlone{}~\cite{Mardirossian2014/10.1039/C3CP54374A}, a range-separated hybrid functional based on the Becke-97 GGA functional~\cite{Becke1997/10.1063/1.475007} and the VV10 non-local correlation functional~\cite{Vydrov2010/10.1063/1.3521275}.
\BninetyVlone{} has been proven to be very reliable across a wide range of benchmarks and shows good agreement with ``gold-standard'' methods on non-covalent interactions of water molecules~\cite{Mardirossian2014/10.1063/1.4868117, Brauer2016/10.1039/C6CP00688D, Manna2017/10.1021/acs.jctc.6b01046}.
The first post-HF method to be used for bulk liquid water was MP2~\cite{DelBen2013/10.1021/jz401931f}.
It uses the exact HF exchange, corrected by second order perturbation theory to account for electron correlation effects.
Because the computational cost scales as $O(N^5)$ with the number of electrons $N$, the use of MP2 in bulk systems remains limited.
Therefore, simulations of bulk water using MP2 have either used a fragment-based approach~\cite{Willow2015/10.1038/srep14358, Liu2017/10.1039/C7CP00667E, Liu2022/10.1021/acs.jpca.2c00601} or are constrained in the size of the basis set, even when the simulations are aided by MLPs~\cite{DelBen2013/10.1021/jz401931f, Li2025/10.4208/cicc.2025.88.01}.

When it comes to the benchmarking of methods to reproduce physical observables, most of the attention has been focused on the electronic structure method, dispersion interactions, and NQEs.
At the same time, some of the underlying settings of the calculation have not been given sufficient consideration, even though some studies reported relevant differences in RDFs, densities, and diffusion through changes in the electronic structure code, the basis set, or changes in the specifics of the dispersion correction~\cite{Lee2007/10.1063/1.2718521, Wang2011/10.1063/1.3521268, DelBen2013/10.1021/jz401931f, DelBen2015/10.1063/1.4927325, Galib2017/10.1063/1.4986284, deHijes2024/10.1063/5.0227514, Dao2026/10.26434/chemrxiv.15000644/v1, ONeill2026/10.48550/arXiv.2605.28798}.
Galib et al.~\cite{Galib2017/10.1063/1.4986284} compared water simulated with \revPBElone{} for different basis sets and found differences in the RDF that were attributed to the parametrization of D3.
Del Ben et al.~\cite{DelBen2013/10.1021/jz401931f, DelBen2015/10.1063/1.4927325} found that including the auxiliary density matrix method (ADMM) approximation in \mbox{PBE0-D3} simulations leads to a slightly less dense and more structured water.
Montero de Hijes et al.~\cite{deHijes2024/10.1063/5.0227514} found that the damping function employed in the D3 correction of \revPBElone{} can impact the temperature--density curve considerably.
These underlying parameters play an important role for ab initio MD and will continue to play an important role as MLPs become more and more widespread.
However, when ab initio calculations are used to drive MD, basis sets, potentials, and convergence settings have to accommodate the existing constraints in computational resources.
Yet, when using the ab initio calculations as reference data to train MLPs, usually the already established setups were used once again without investigating other options.
Especially all-electron potentials have rarely been explored for production-level MD simulations of bulk liquid systems using MLPs.
Additionally, recent work has shown that many popular databases used to develop MLPs contain a high number of structures with non-zero net forces due to insufficient basis set cutoffs and self-consistent field (SCF) convergence settings~\cite{Kuryla2025/10.48550/arXiv.2510.19774}.
The advent of MLPs opens the door to routine MD simulations based on high-accuracy methods evaluated with precision, as MLPs' computational efficiency finally allows us to ignore the fine balance between cost and accuracy and fully commit to obtaining the true PES associated with a certain electronic structure method.
Combined with efficient construction of training sets, MLPs can also be used to systematically screen DFT functionals and electronic wave function methods close to the convergence limit based on calculations of macroscopic properties.
In this work, we investigate if common choices for basis sets and potentials are sufficiently converged.
We do this by training four committees of neural network potentials (C-NNPs)~\cite{Schran2020/10.1063/5.0016004} on different electronic structure setups and calculating physical observables from MD trajectories obtained with these C-NNPs.
Our baseline is a common setup for \revPBElone{}, which is known for an impressive agreement with experimental reference data.
We compare this setup with one that uses the same functional and dispersion correction but with a highly converged basis set and an all-electron potential, and one that uses the same tight settings in combination with a \Bninetylone{} functional.
The fourth model uses a common setup for MP2, a method known for being sensitive to basis set completeness.
We examine the role of different convergence settings in the validation of these MLPs.
Subsequently, we calculate the radial distribution function (RDF), pressure--density curve, diffusion coefficient, and hydrogen bond lifetime and analyze covalent and hydrogen bonds in detail.
Where available, we compare these results with experimental findings.
With this meticulous testing scheme, we interrogate whether the impressive agreement of \revPBElone{} with experiment is due to the excellent performance of the functional or whether some fortuitous cancellation of errors plays a critical role.
Our classical and quantum simulations reveal how the changes in the PES induced by practically converged settings play out for thermally averaged static and dynamic properties.

\section{Computational Details}
\label{sec:comp-details}

The electronic structure calculations for training and test sets are executed using the open-source package CP2K~\cite{Kuehne2020/10.1063/5.0007045} utilizing the Quickstep framework~\cite{Vandevondele2005/10.1016/j.cpc.2004.12.014}.
In total, four different main setups are used for the calculations, with each of them including dispersion interactions in some form.
As our baseline method, we use the revPBE0 hybrid density functional~\cite{Perdew1996/10.1103/PhysRevLett.77.3865, Zhang1998/10.1103/PhysRevLett.80.890, Adamo1999/10.1063/1.478522} combined with Grimme's DFT-D3 dispersion correction with zero damping~\cite{Grimme2010/10.1063/1.3382344, Goerigk2011/10.1039/c0cp02984j},
a setup identical to our previous work~\cite{Marsalek2017/10.1021/acs.jpclett.7b00391, Schran2020/10.1063/5.0016004}.
These calculations use the Gaussian and plane waves method~\cite{Lippert1997/10.1080/002689797170220} (GPW) in combination with the GTH-PBE pseudopotentials~\cite{Goedecker1996/10.1103/PhysRevB.54.1703, Krack2005/10.1007/s00214-005-0655-y}.
We employ a robust implementation of HF exchange in periodic systems~\cite{Guidon2009/10.1021/ct900494g}, supplementing our primary TZV2P Gaussian basis set with the cpFIT3 auxiliary basis set for the auxiliary density matrix method (ADMM)~\cite{Guidon2010/10.1021/ct1002225}.
The GPW plane-wave cutoff is \SI{400}{Ry}
and the SCF convergence threshold is set to $5 \times 10^{-7}$.
We will refer to these settings as \revPBET.
For the \revPBEQ{} setup, we keep the functional, but tighten several key aspects of the calculations.
First, we remove the pseudopotentials and perform an all-electron calculation using the Gaussian and augmented plane waves~\cite{Lippert1999/10.1007/s002140050523, Krack2000/10.1039/b001167n} (GAPW) method instead of GPW.
We get close to the complete basis set limit by changing the primary basis set to def2-QZVP and the ADMM auxiliary basis set to def2-TZVP, both introduced by Alrichs~\cite{Ahlrichs2005/10.1039/B508541A}.
Furthermore, we increase the plane-wave cutoff to \SI{800}{Ry} and tighten the SCF convergence threshold by two orders of magnitude to $5 \times 10^{-9}$.
To better analyze the separate roles that the potential and basis set play, we also create a \revPBEQGTH{} setup that is identical to \revPBEQ{}, but uses the GTH pseudopotentials instead of all-electron potentials.
For the \Bninety{} setup, we keep these improved convergence settings, but replace the revPBE0-D3 functional with the \Bninetylone{}~\cite{Mardirossian2014/10.1039/C3CP54374A} range-separated hybrid functional.
To include dispersion interactions, it uses the revised VV10 non-local correlation functional~\cite{Sabatini2013/10.1103/PhysRevB.87.041108, Mardirossian2017/10.1021/acs.jpclett.6b02527} (indicated by the ``-rV''), an adaptation of the VV10 functional~\cite{Vydrov2010/10.1063/1.3521275} optimized for evaluation in plane wave basis sets, with parameters $b$=6.0 and $C$=0.01.

The final setup, denoted \MPtwo{}, uses second order Møller–Plesset perturbation theory~\cite{Moller1934/10.1103/PhysRev.46.618} to treat electronic correlation explicitly.
Unfortunately, running all-electron calculations of bulk structures close to the basis set limit for MP2 would be prohibitively expensive for our geometries, even for single-point calculations.
Hence, we use the Resolution of Identity Gaussian and Plane Wave~\cite{DelBen2012/10.1021/ct300531w, DelBen2013/10.1021/ct4002202, DelBen2015/10.1063/1.4919238} (RI-GPW) approach with GTH pseudopotentials optimized for HF theory in combination with a correlation-consistent triple-$\zeta$ (cc-TZ) basis set~\cite{Vahtras1993/10.1016/0009-2614-93-89151-7, DelBen2013/10.1021/ct4002202} and an auxiliary triple-$\zeta$ basis set for the resolution of identity calculations~\cite{Weigend1998/10.1016/S0009-2614-98-00862-8, DelBen2013/10.1021/ct4002202}.
We set the general plane-wave cutoff to \SI{800}{Ry} and the cutoff for calculating MP2-integrals to \SI{300}{Ry}.
The SCF convergence criterion is set to $10^{-6}$.
This setup has also been used by the first studies of bulk liquid water with MP2~\cite{DelBen2013/10.1021/jz401931f} as well as one study using MLPs~\cite{Li2025/10.4208/cicc.2025.88.01}, which will become useful to compare results.

As an example of computational demands of the different methods, we can examine the run time and peak memory utilization with the different setups for 64 bulk water molecules.
To accelerate SCF convergence, we use converged Kohn--Sham orbitals from the GGA counterpart to initialize each hybrid functional's SCF procedure, and PBE orbitals to initialize the HF SCF in the RI-MP2 calculation.
A single-point calculation with the original \revPBET{} method on a node with 2 AMD EPYC 7301 16-core CPUs took 4 minutes and required \SI{18}{\giga\byte} of memory.
A single-point calculation on the same hardware took roughly 14 minutes for \revPBEQ{} and 34 minutes for \Bninety{} and required \SI{35}{\giga\byte} and \SI{123}{\giga\byte} of memory, respectively.
The \MPtwo{} calculations were run on 16 nodes with 2 AMD EPYC 7H12 64-core CPUs each, and a single point calculation took circa 28 minutes with a memory requirement of \SI{3.2}{TB}.
Further details on the resource requirements can be found in the Supporting information \ref{si-sec:resources}.

The training data comprises a total of 964 structures from different aqueous systems, which were selected from MD trajectories using the query by committee (QbC) process~\cite{Schran2020/10.1063/5.0016004}.
We extended the 814-structure canonical ensemble ($NVT$) water dataset by Schran et al.~\cite{Schran2020/10.1063/5.0016004} with 150 additional structures sampled from MD simulations in the isobaric--isothermal ensemble ($NpT$) to improve the density prediction of the model.
Further details can be found in Section~\ref{si-sec:training_data}.
The data set now contains bulk structures of liquid water (64 molecules), ice \ih (96 molecules), ice VIII (64 molecules), as well as a water slab (216 molecules), taken from classical and path integral MD (PIMD) simulations at different temperatures and densities.
Energies and forces are evaluated for these geometries at the different levels of electronic structure theory that we introduced above to create the four training sets.
Unfortunately, the memory requirements of a 216-molecule slab surpass our available resources, so the MP2 training set is limited to liquid water and bulk ice structures.

The C-NNPs consists of 8 separate Behler--Parrinello deep neural networks~\cite{Behler2007/10.1103/PhysRevLett.98.146401, Behler2011/10.1063/1.3553717} as implemented in the N2P2 package~\cite{Singraber2019/10.1021/acs.jctc.8b00770, Singraber2019/10.1021/acs.jctc.8b01092}.
Each committee member is trained on a different 90\% subset of the total training data set and initialized with different weights following the Nguyen--Widrow scheme~\cite{Nguyen1990/10.1109/IJCNN.1990.137819} at the start of each training procedure.
The mean over all committee members forms the eventual PES prediction.
The NNPs are trained on both energies and forces using a multi-stream adaptive extended Kalman filter~\cite{Shah1992/10.1016/S0893-6080(05)80139-X, Blank1994/10.1002/cem.1180080605}, and the remaining training settings follow the recommendations for water from the N2P2 developers.
The NNPs are trained for 1000 epochs, substantially longer than in previous work, as our tests have shown no signs of overtraining even for such a high number of epochs but instead a slight improvement of predictions on an independent test set.
As input features, we use atom-centered symmetry functions~\cite{Behler2011/10.1063/1.3553717} (ACSF) with 32 pairwise functions (8 functions for each combination of hydrogen and oxygen atoms) and 25 angular symmetry functions~\cite{Morawietz2016/10.1073/pnas.1602375113}.
Before training, each feature is pre-processed by subtracting its mean across the whole data set and scaling the values to cover a range of 1.

With these four C-NNPs, we run classical and path integral MD and simulations~\cite{Markland2018/10.1038/s41570-017-0109} in both the $NVT$ and $NpT$ ensemble.
These simulations are performed using the i-PI package~\cite{Kapil2019/10.1016/J.CPC.2018.09.020} with interactions provided by CP2K~\cite{Kuehne2020/10.1063/5.0007045}, which has an efficient implementation of committees of Behler--Parrinello potentials~\cite{Schran2020/10.1063/5.0016004}.
The simulation boxes for liquid water contain 512 molecules, with a \SI{24.84}{\angstrom} cubic box in $NVT$ calculations, corresponding to the experimental density of liquid water at ambient conditions.
For ice \ih, we use a system of 96 molecules in a box of dimensions 13.489$\times$15.576$\times$14.641 \si{\angstrom}, again in accordance with experimental results.
The temperatures are set to \SI{300}{K} and \SI{250}{K} for liquid water and ice \ih, respectively.
The production $NVT$ trajectories of bulk liquid water have a length of \SI{1}{ns} with a \SI{0.5}{fs} time step for classical MD, and a length of \SI{250}{\pico\second} with a \SI{0.25}{fs} time step for PIMD.
The $NpT$ simulations have a length of \SI{100}{ps} at each pressure.
In classical MD simulations, the temperature is controlled by a velocity rescaling thermostat~\cite{Bussi2007/10.1063/1.2408420} with a time constant $\tau$ of \SI{1}{ps}.
For the path integral simulations, we perform thermostatted ring polymer molecular dynamics (TRPMD)~\cite{Craig2004/10.1063/1.1777575, Rossi2014/10.1063/1.4883861} with 32 replicas, employing a global path integral Langevin equation (PILE-G) thermostat~\cite{Ceriotti2010/10.1063/1.3489925} with $\tau$ = \SI{1}{ps} and $\lambda_\mathrm{PILE}$ = 0.5.
The $NpT$ simulations are run in an isotropically scaling cell at pressures of 1, 1000, 2000, 3000, 4000, and 5000 \si{atm} for classical simulations and 1, 2000 and 5000 \si{atm} for PIMD.
The barostat is a Bussi--Zykova--Parrinello barostat~\cite{Bussi2009/10.1063/1.3073889} with a time constant of \SI{200}{fs}, thermostatted by a smart-sampling generalized Langevin equation (GLE) thermostat~\cite{Ceriotti2009/10.1103/PhysRevLett.102.020601, Ceriotti2010/10.1021/ct900563s}.

Moving on to dynamical properties, the diffusion coefficient $D$ can be expressed in terms of the mean square deviation $\mathrm{MSD}$ using Einstein’s relation:
\begin{align}
    D =
    \frac{\mathrm{MSD}}{6t} =
    \frac{1}{N} \sum_i \frac{\left \langle \left| \vect{r}_i(0) - \vect{r}_i(t) \right|^2 \right \rangle}{6t},
\end{align}
where $\vect{r}_i(t)$ are the positions of $N$ atoms at time $t$.
To correct for finite-size effects in a periodic cell, we add the correction term~\cite{Dunweg1993/10.1063/1.465445}
\begin{align}
    D_\mathrm{corr} = \frac{k_\mathrm{B} T\xi}{6\pi\eta} \frac{1}{L},
\end{align}
where $k_\mathrm{B}$ is the Boltzmann constant, $T$ is the temperature, $\eta$ is the shear viscosity of the liquid, and $\xi=2.83729$ accounts for the cubic shape of the cell of length $L$.
We use the experimental value $\eta=0.8925$~\si{\milli\pascal\second} for water at 300~\si{\kelvin} for all our models.
This is only an estimate of the true correction factor, which would use the shear viscosity of each model.
As an alternative, extrapolation to an infinite cell size based on simulations at multiple different system sizes could also be used~\cite{Morawietz2016/10.1073/pnas.1602375113}.

The hydrogen bond lifetimes are calculated from the hydrogen bond existence criteria $h_n(t)$
\begin{align}
    h_n(t) = \Theta(r_n(t) - r_0)\, \Theta(\theta_n(t) - \theta_0),
\end{align}
where $\Theta$ is the step function and $n$ is an index that runs over all potential hydrogen bonds in the system.
$r_n(t)$ is the distance between the two oxygen atoms and $\theta_n(t)$ is the angle between the vector connecting the two oxygen atoms and the vector connecting the donor oxygen and its covalently bonded hydrogen atom at time $t$.
We use the cutoffs $r_0=3.5$ \si{\angstrom} and $\theta_0=30\,\degree$ for the distance and angle, respectively~\cite{Luzar1993/10.1063/1.464521, Luzar1996/10.1038/379055a0}.
For path integral trajectories, we use the RPMD average over the $P$ replicas and thus have
\begin{align}
    h_n(t) = \frac{1}{P} \sum^P_{j=1} \Theta(r_{n}^{(j)}(t) - r_0)\, \Theta(\theta_{n}^{(j)}(t) - \theta_0),
\end{align}
where $r_{n}^{(j)}$ and $\theta_{n}^{(j)}$ are the distance and angle for replica $j$.
This $h_n(t)$ can now take fractional values between 0 and 1 with a denominator $P$.
We calculate the autocorrelation function of $h_n(t)$ averaged over all $N_{\mathrm{HB}}$ hydrogen bonds that occur in the trajectory as
\begin{align}
    C(\tau) =
    \frac{1}{N_\mathrm{HB}} \sum_{n=1}^{N_\mathrm{HB}} \frac{\left< h_n(t_0)\, h_n(t_0 + \tau) \right>_{t_0}}{\left<h_n\right>}
    ,
\end{align}
which gives the probability that a hydrogen bond that existed at time $t_0$ exists at time $t_0 + \tau$~\cite{Rapaport1983/10.1080/00268978300102931, Luzar1996/10.1038/379055a0}.
Finally, we integrate $C(\tau)$ over $\tau$ to obtain a time that characterizes the lifetime of the hydrogen bond.

\section{Results}

We will start by inspecting the four C-NNPs and validating that all of them reliably reproduce the PES of their reference method.
Next, we will calculate physical observables from the trajectories obtained with these models.
By comparing them with each other and with experimental reference data, we will elucidate the role that pseudopotentials, basis sets, and exchange--correlation functionals play in the simulation of water.

\subsection{Model validation}

\begin{figure}
\centering
\includegraphics{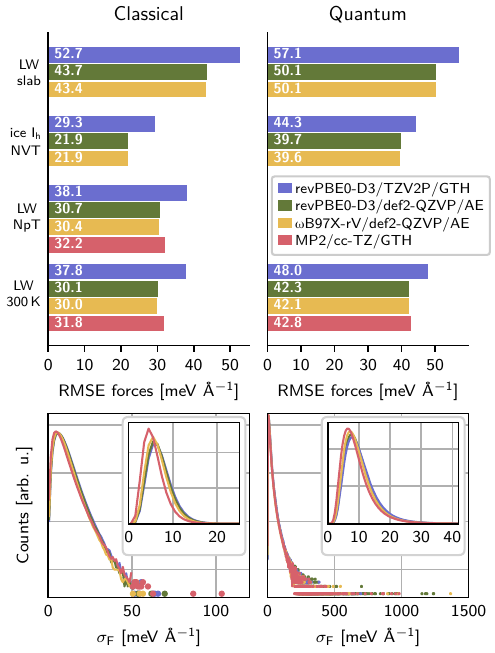}
\caption{\label{fig:testset}
    Top panels: RMSEs for the 4 different C-NNPs on test sets of liquid water, ice and a slab with a liquid-vacuum interface for both classical (left) and quantum (right) structures.
    Bottom panels: Distribution of force disagreements along an $NVT$ trajectory of liquid water using classical (left) and path integral (right) MD.
    The main panels use a logarithmic scale for the y-axis, while the insets shows the distribution for low disagreements on a linear scale.
}
\end{figure}

Comparing the predictions of a model to a reference dataset independent from the training dataset is considered the gold standard for validating machine learning models.
For this purpose, we use the test set introduced by Schran et al.~\cite{Schran2020/10.1063/5.0016004}
The test set was initially constructed to validate the performance of the C-NNP in comparison to ab initio calculation with the \revPBET{} setup.
We made some modifications to address the increased computational demands of our highly converged setups and MP2 calculations, as well as the inclusion of $NpT$ simulations.
The test set consists of 500 structures sampled from each classical MD and PIMD simulations in the $NVT$ ensemble of bulk liquid water and ice \ih, as well as 200 structures from each classical and path integral MD of a slab of liquid water.
Since we will use the MLPs to calculate pressure--density curves, we also added structures sampled from $NpT$ simulations.
For this test set, we sampled 100 structures each from $NpT$ simulations at 1, 1000, 2000, 3500, and 5000 atm to obtain a total of 500 structures.
For the MP2 test set, it was necessary to reduce the number of structures to 100 each from classical and quantum simulations of $NVT$ and $NpT$ trajectories at \SI{300}{K}.
Furthermore, no ice structures are included in the MP2 test set, as these have been shown to score the best out of all systems for every model.
Therefore, we do not expect negative outliers in this phase.
The force root mean square errors (RMSEs) of a selection of these test sets are shown in the top panels of Figure~\ref{fig:testset}, while the remaining test sets as well as the energy errors can be found in Figure~\ref{si-fig:testsets}.
The errors for all models across all test sets are low enough to reliably reproduce the static and dynamic properties of the ab initio reference methods.~\cite{Cheng2019/10.1073/pnas.1815117116, Schran2020/10.1063/5.0016004, Schran2021/10.1073/PNAS.2110077118}
As expected, the errors for ice are lower than for liquid water due to the narrower exploration of the configuration space.
Meanwhile, the errors for the slab test sets are higher because the molecules at the liquid--vapor interface are more challenging for the model, and there is also less reference data for them in the training set.
Surprisingly, the errors of the models trained on the recalculated data are even lower than for the original model.
This, however, is not primarily a result of more accurate predictions, but of inaccuracies in the \revPBET{} test sets.
Due to the so-called ``egg box effect''~\cite{Durham2025/10.1088/2516-1075/adc056}, which is a major source of error in the GPW calculations of \revPBET{}, but suppressed in the GAPW and MP2 calculations, all \revPBET{} datasets suffer from a small amount of additional noise.
During training, the NNPs are not capable of learning the egg box effect, which depends on absolute atomic positions, because the ACSFs depend only on relative atomic positions.
Therefore, the egg box effect adds aleatoric uncertainty to the training data.
The models will smooth over most of this added noise, but the egg-box variation in test-set reference calculations contributes to the error when evaluating the performance of a model.
Recently, Kuryla et al.~\cite{Kuryla2025/10.48550/arXiv.2510.19774} made similar observations when they discovered a high number of structures with non-zero net forces in many popular databases used to develop MLPs.
These net forces indicate errors in the reference calculations caused by inadequate SCF settings and insufficient basis sets.
They defined a net force above \SI{1}{meV/\angstrom/atom} as the threshold for problematic structures.
We found substantial net forces for our \revPBET{} dataset --- an average of \SI{2.14}{\milli\eV\per\angstrom\per atom} over all its structures.
For the remaining datasets, the net forces were significantly lower --- \SI{0.08}{\milli\eV\per\angstrom\per atom} for both \revPBEQ{} and \Bninety{} and \SI{0.002}{\milli\eV\per\angstrom\per atom} for \MPtwo.
Having identified this issue, Kuryla et al. attempted to correct calculation settings for many popular electronic structure codes.
However, for CP2K, a considerable non-zero net force still remained after tightening many key settings.
We show that switching from the GPW to the GAPW method helps reduce the remaining net forces considerably.
We recalculated the ``LW NpT'' set with the \revPBET{} settings, but using the GAPW method and a higher plane wave cutoff of \SI{800}{Ry} and obtained net forces of \SI{0.07}{meV/\angstrom/atom} and test set errors similar to the other GAPW methods.
This shows that GAPW's suppression of the egg-box effect~\cite{Lippert1999/10.1007/s002140050523} is an important tool for reducing inconsistencies in CP2K calculations.
Further insights into this issue and its effect on training, including for modern high-capacity models, can be found in Section~\ref{si-sec:egg-box}.

Since the different reference methods also change the structures sampled by MD, we compared the distribution of committee disagreement in MD simulations using the established \revPBET{} C-NNP to those using the models trained on the new reference methods.
The distributions are displayed in the bottom panels of Figure~\ref{fig:testset}.
They are very similar for all four C-NNPs, indicating reliable predictions.
For path integral simulations, we find a number of structures with noticeably high ($>$ \SI{500}{\milli\eV\per\angstrom}) force disagreements.
However, as the inset in the bottom right panel emphasizes, these cases represent a negligible fraction of the total data and therefore do not negatively impact the final results.
These committee disagreements in combination with the low test set errors prove that recalculating energies and forces for the geometries of the original training set is a viable strategy to obtain a reliable C-NNP for a different reference method.
In principle, the model for the new method could show an appreciable increase in disagreements during MD or in errors evaluated on a test set.
In that case, one could simply continue the QbC-based active learning that created the original training set to include additional structures required to make the new PES well-determined by the data.

\subsection{MD Simulations}

We will now turn our focus to analyzing the MD trajectories.
With the exception of the pressure--density analysis, we always use $NVT$ simulations of bulk liquid water and keep a consistent color scheme for all plots:
\revPBET{} is shown in blue, \revPBEQ{} in green, \Bninety{} in yellow, and \MPtwo{} in red.
Results are shown as solid lines/bars for classical MD and as dashed lines/bars for PIMD.

\begin{figure}
\centering
\includegraphics{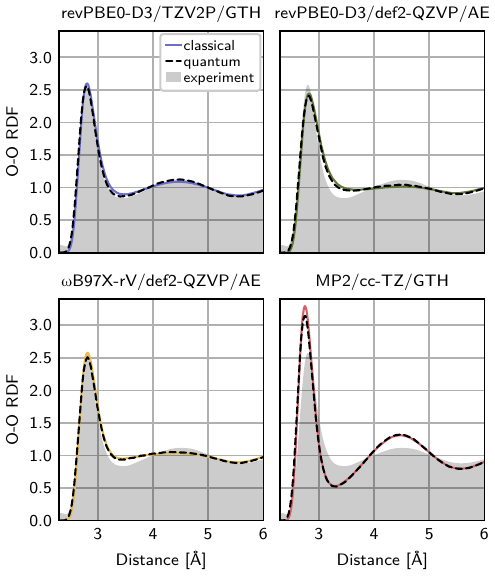}
\caption{\label{fig:rdf}
    The oxygen-oxygen RDF for the 4 C-NNPs.
    In each panel, the result from the classical MD simulation is shown by the solid line in the respective color, the PIMD result is shown by the black dashed line and the experimental reference~\cite{Skinner2013/10.1063/1.4790861} is indicated by the gray-shaded area.
}
\end{figure}

We start the investigation of structural properties of liquid water by examining the oxygen--oxygen (O--O) radial distribution function (RDF) shown in Figure~\ref{fig:rdf}.
It has been shown in previous studies~\cite{Marsalek2017/10.1021/acs.jpclett.7b00391} that classical \revPBET{} matches the experimental results~\cite{Skinner2013/10.1063/1.4790861} closely.
Adding NQEs improves the result even further by slightly shortening the onset of the first coordination shell and by deepening the the first minimum.
Unfortunately, the fully converged \revPBEQ{} setup shows that this immaculate match to the experimental benchmark is in part enabled by the incomplete convergence of the original computational setup.
The converged RDF features a flatter peak at the first hydration shell and a mostly flat RDF beyond that.
As can be seen in Figure~\ref{si-fig:rdf_oo}, the main difference in the RDF stems from the change in basis set, whereas the all-electron potential contributes only negligibly.
It should be noted that in addition to this highly converged primary basis set, we also switch to a considerably richer ADMM auxiliary basis set.
Del Ben et al. suggests in a study using PBE0 that a small ADMM basis set can lead to an increased structure in the O--O RDF~\cite{DelBen2013/10.1021/jz401931f, DelBen2015/10.1063/1.4927325}.
NQEs slightly improve the results, but the difference from the experimental findings is still substantial.
Galib et al. have run calculations for \mbox{revPBE}, the GGA counterpart of \mbox{revPBE0}, with the TZV2P and MOLOPT-DZVP-SR-GTH basis sets and found that without dispersion correction the O--O RDFs for both setups matched well, but including D3 dispersion corrections lead to substantial differences~\cite{Galib2017/10.1063/1.4986284}.
They attributed this result to the D3 parametrization and its basis set.
\Bninety{} reproduces the key features of the experimental results moderately better, though a considerable error remains.
A study by Yao et al. used \Bninetylone{} with the TZV2P basis set and GTH pseudopotentials and reported an overstructured O--O RDF~\cite{Yao2018/10.1021/acs.jctc.7b00846}.
This confirms the trend of a larger basis set loosening the water structure we observed for \revPBElone{}.
While the RDFs of all three DFT sets  clearly resemble the experimental one, \MPtwo{} is a strong outlier.
Overall, the RDF is severely over-structured, with the peak of the first coordination shell being way too high and at a too short distance.
Subsequent peaks are too high and valleys are too deep.
These findings for MP2 agree well with the original ab initio MD findings~\cite{DelBen2013/10.1021/jz401931f, DelBen2015/10.1063/1.4927325} and the more recent MLP simulations using the same setup for the electronic structure reference calculations~\cite{Li2025/10.4208/cicc.2025.88.01}.
They attributed this behavior to an overestimation of binding energies due to the small basis set.
Other studies using a fragment-based MP2 approach in combination with a high-quality basis set obtained a better fitting RDF~\cite{Willow2015/10.1038/srep14358, Liu2017/10.1039/C7CP00667E, Liu2022/10.1021/acs.jpca.2c00601}.
D\`ao et al.~\cite{Dao2026/10.26434/chemrxiv.15000644/v1} used the cc-TZ basis in combination with \revPBElone{} and found that it leads to an overstructuring of the RDF.
Although we should not draw any direct conclusions from these hybrid DFT results for MP2 calculations using the same basis set, it further emphasizes how the basis set impacts the structural properties.

\begin{figure}
\centering
\includegraphics{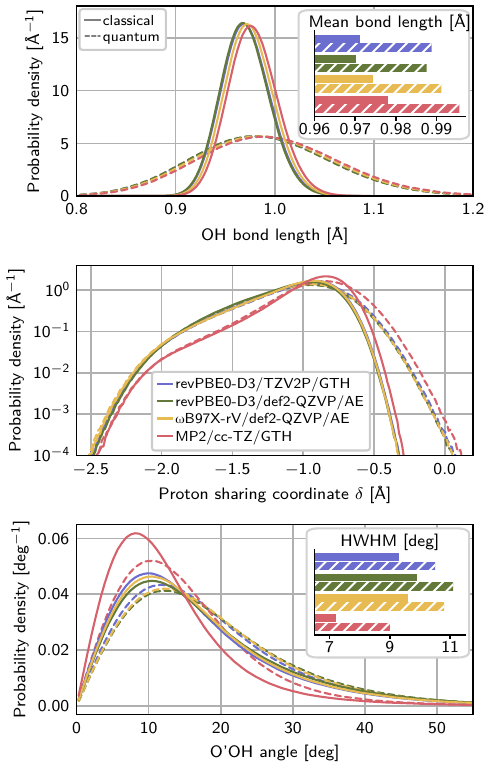}
\caption{\label{fig:bonds_angles}
    Analysis of the structure of covalent and hydrogen bonds.
    Top panel: The distribution and means of lengths of covalent oxygen-hydrogen bonds for classical (solid lines/bars) and PI (dashed lines/bars) MD run with the 4 different C-NNPs.
    The main plot shows the distributions of bond lengths and the inset the mean bond lengths.
    Middle panel: The distributions of the proton sharing coordinate.
    Bottom panel: The distributions of the hydrogen bond angle.
    The inset compares the half width at half maximum of the different methods.
}
\end{figure}

For further insights, we will look closer at the structure of the water molecules and the interactions with the first coordination shell.
As shown in the top panel of Figure~\ref{fig:bonds_angles} and the inset bar chart, the changes in the reference method have only a small effect on the lengths of the covalent bonds.
For classical MD, the means of all 4 distributions lie within less than \SI{0.01}{\text{\AA}}, or less than \SI{1}{\%} of the total bond lengths, with \revPBEQ{} having the shortest covalent bonds and \MPtwo{} the longest.
NQEs lead to much wider distributions, whose means are moderately longer than those for classical nuclei.
The order of the ab initio methods remains the same and the difference between them remains almost perfectly constant.
Tests with the \revPBEQGTH{} setup (see Figure~\ref{si-fig:bonds_angles}) suggest that the changes in basis set and potential result in opposite outcomes, effectively canceling each other out.
The middle panel of Figure~\ref{fig:bonds_angles} shows the proton sharing coordinate $\delta = d_{\mathrm{HO}} - d_{\mathrm{HO^\prime}}$, which measures the difference of the distances between the hydrogen atom and the donor and acceptor oxygen atoms.
Quantum delocalization shifts the distribution towards protons being shared more equally between the two oxygen atoms.
In rare cases (\SI{0.0015}{\%} to \SI{0.0045}{\%}), this even results in the so-called ``proton excursion'', where the proton is closer to the acceptor atom than the donor atom, i.e., $\delta > 0$.
None such cases can be observed without NQEs.
These numbers for \revPBET{} agree well with previous findings~\cite{Marsalek2017/10.1021/acs.jpclett.7b00391}.
Comparing the methods, a split between the DFT functionals and MP2 is apparent.
The MP2 trajectories exhibit a shift towards more shared protons, as well as flatter distributions at low $\delta$ and more pronounced peaks.
The bottom panel of Figure~\ref{fig:bonds_angles} shows the angle between the donor--acceptor vector and the covalent bond vector.
NQEs generally shift the distributions towards wider angles compared to their classical counterparts, while the distribution becomes flatter.
When comparing the DFT setups, we see that the highly-converged \revPBElone{} shows a wider distribution compared to the standard settings, with \Bninetylone{} being close to \revPBEQ{}.
The half width at half maximum (HWHM) of the distributions serves as a good measure for fluctuations in the HB angle, which are related to the strength of the hydrogen bonds~\cite{Morawietz2016/10.1073/pnas.1602375113} --- a broader distribution corresponds to weaker hydrogen bonds.
Furthermore, for all our models, we find consistently smaller HWHM than Morawietz et al., who employed GGA functionals with and without dispersion corrections~\cite{Morawietz2016/10.1073/pnas.1602375113}.
As with the $\delta$ coordinate, the differences between MP2 and the DFT functionals are considerable.
The MP2 peaks are shifted towards sharper angles and the tail of the distribution above \SI{25}{\degree} is strongly suppressed.

\begin{figure}
\centering
\includegraphics{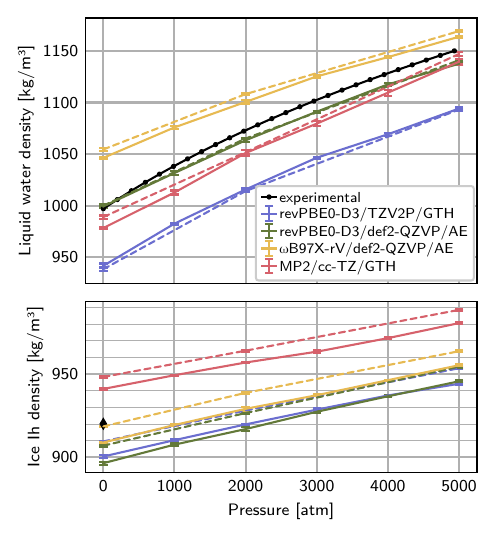}
\caption{\label{fig:pressure-density}
    The pressure--density curve of liquid water (top) and ice \ih{} (bottom) for the 4 different C-NNPs.
    The black curve in the top panel and the black diamond in the bottom panel shows the experimental reference for each system~\cite{Grindley1971/10.1063/1.1675455, Sanz2004/10.1103/PhysRevLett.92.255701}.
    The error bars indicate statistical errors obtained by block averaging.
    As in the other plots, solid lines are used for classical MD and dashed lines for PIMD.
}
\end{figure}

One of the most intensely debated topics in ab initio and MLP water simulations is the density~\cite{Schmidt2009/10.1021/jp901990u, Habershon2009/10.1063/1.3167790, Wang2011/10.1063/1.3521268, Ma2012/10.1063/1.4736712, Morawietz2016/10.1073/pnas.1602375113, Galib2017/10.1063/1.4986284, deHijes2024/10.1063/5.0227514, deHijes2024/10.1063/5.0197105}.
Therefore, we analyze $NpT$ trajectories and test how well the different methods can reproduce the density at different pressures.
Figure~\ref{fig:pressure-density} shows the pressure--density curve for water (top panel) and ice \ih{} (bottom panel) for pressures between 1 and \SI{5000}{atm} and compares them with experimental results~\cite{Grindley1971/10.1063/1.1675455} (indicated in black).
It shows how the established setup \revPBET{} underestimates the density of both liquid water and ice.
Our results agree well with a study by Cheng et al., who used an MLP trained on the same reference method but with a training set constructed in a way different from ours~\cite{Cheng2019/10.1073/pnas.1815117116}.
Upgrading the basis set has a considerable impact on the density of liquid water, giving us a curve that closely matches the experimental curve, especially for low densities.
The high sensitivity of the pressure to the basis set has been remarked previously in the context of plane wave basis sets~\cite{deHijes2024/10.1063/5.0197105}.
In our case, the improvement comes from the increased flexibility of the primary Gaussian basis set used for the Kohn--Sham orbitals.
The density given by these orbitals then needs to be expanded in the auxiliary basis set, which in the case of the GAPW method is a combination of plane waves and local corrections based on the primitive Gaussians of the primary basis set.
On the other hand, tests with pseudopotentials presented in Figure~\ref{si-fig:density} show that the impact of the potential on the density is negligible.
Although the density does not increase fast enough with increasing pressure, it is still in very good agreement with the experimental results.
Willow et al. showed in tests with parametrizable force fields that polarizability is an important factor in determining at which pressure water has a density of \SI{1.000}{\gram\per\centi\meter\cubed}~\cite{Willow2016/10.1021/acs.jpclett.5b02430}.
A higher polarisability, which is usually associated with larger basis sets~\cite{Rappoport2010/10.1063/1.3484283, Willow2016/10.1021/acs.jpclett.5b02430, Afzal2019/10.1039/C8CP05492D, Kumar2025/10.1002/hlca.202400130}, corresponds to a lower pressure at which experimental ambient density is reached~\cite{Willow2016/10.1021/acs.jpclett.5b02430}.
The ADMM basis set and the interplay between dispersion correction and basis set, both of which we have already discussed for the RDF, likely play a role here as well~\cite{DelBen2013/10.1021/jz401931f, Galib2017/10.1063/1.4986284}.
A study using \revPBElone{} in a highly-converged plane wave setup obtained an even higher density of roughly \SI{1.03}{\gram\per\centi\meter\cubed}~\cite{deHijes2024/10.1063/5.0227514}.
For ice \ih, however, both \revPBElone{} models give almost identical results.
\Bninety{} overestimates the density of liquid water across the whole range of pressures, although more strongly for low pressures than for high pressures.
For ice \ih, it reproduces the experimental reference point~\cite{Sanz2004/10.1103/PhysRevLett.92.255701} the best out of all the models, especially when including NQEs.
Despite the unsatisfactory RDF, \MPtwo{} gives a decent pressure density curve for liquid water.
Even though it is about \SI{0.02}{\gram\per\centi\meter\cubed} too low, it replicates the overall shape of the experimental result better than our other setups.
It is also in decent agreement with ab initio results, considering their limited number of Monte Carlo cycles~\cite{DelBen2013/10.1021/jz401931f}.
On the other hand, MP2 overestimates the experimental density for ice \ih.
As for all models, the density of ice increases linearly with pressure while being lower than that of liquid water.
When comparing classical and path integral simulations of liquid water and ice, we see an increase in density between 0.005 and 0.011 \si{\gram\per\centi\meter\cubed} in most cases when including NQEs.
The two exceptions are liquid water with the \revPBElone{} models, where we observe no NQEs on the density outside the error bars.
This is in contrast to a density increase due to NQEs reported in the above-mentioned Ref.~\citenum{Cheng2019/10.1073/pnas.1815117116}.

\begin{figure}
\centering
\includegraphics{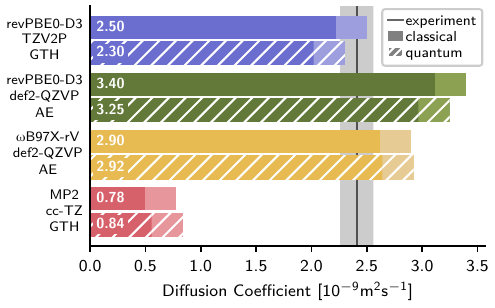}
\caption{\label{fig:diffusion}
    Comparison of the diffusion coefficients for the different C-NNPs in classical and quantum MD.
    The pale colors at the end of each bar signal the magnitude of the finite-size corrections added to the calculated values.
    The gray vertical area indicates the experimental reference value including its confidence interval~\cite{Holz2000/10.1039/b005319h}.
}
\end{figure}

Let us now take a look at dynamic properties, starting with a comparison of the diffusion coefficient for the different methods.
Figure~\ref{fig:diffusion} shows the diffusion coefficients we obtained and compares them with the experimental value of $(2.41 \pm 0.15)\times10^{-9}$ \si{\meter\squared\per\second}~\cite{Holz2000/10.1039/b005319h}.
The lightly-shaded area of the bars indicates the contribution from the correction we applied to account for the finite size of the simulation cell.
Our results for \revPBET{} match ab initio findings (corrected at a different cell size) adequately~\cite{Marsalek2017/10.1021/acs.jpclett.7b00391}, which further validates the MLPs.
Both classical and path integral simulations give a diffusion coefficient within the confidence interval of the experiment, with a decrease of 8\% due to NQEs.
However, analogously to what we have already seen with the RDF, using a more complete basis set as well as all-electron potentials leads to less structure in the system and increases diffusion by $0.9$ and $0.95\times 10^{-9}$ \si{\meter\squared\per\second} for classical and path integral simulations, respectively.
A comparison of the three \revPBElone{} models in Figure~\ref{si-fig:diffusion} shows that the larger basis set and the all-electron potential both result in faster diffusion, with the basis set having a stronger impact than the potential.
\Bninety{} gives a diffusion coefficient closer to the experimental result, though still somewhat overestimated.
Furthermore, in contrast to \revPBElone{} simulations, including NQEs results only in a negligible increase in diffusion.
Clear problems become apparent for the \MPtwo{} setup, where the diffusion is severely underestimated.
Notably, a substantial part of the total diffusion coefficient does not even originate from actual diffusion in the simulation, but rather from the finite-size correction of $0.28\times10^{-9}$ \si{\meter\squared\per\second} (36\% of the total).
Furthermore, we observe a modest increase in diffusion when NQEs are included.
For comparison, with explicit MP2 AIMD at the classical level, Del Ben et al.~\cite{DelBen2015/10.1063/1.4927325} report $0.67\times 10^{-9}$ \si{\meter\squared\per\second} without finite size corrections in a smaller system of 64 molecules and with much shorter trajectories (2$\times$10 \si{\pico\second}).
MLP simulations by Li et al.~\cite{Li2025/10.4208/cicc.2025.88.01} referencing the same MP2 setup obtained $0.693 \times 10^{-9}$ \si{\meter\squared\per\second}, including a finite size correction based in the experimental viscosity that is a factor of 4 smaller than ours, despite their simulation cell being smaller.
They also report a larger difference between classical and quantum simulations ($1.060 \times 10^{-9}$ \si{\meter\squared\per\second} for PIMD) than our observations.
It is not straightforward to determine how these differences in simulations with what should nominally be the same PES emerge as a result in the specifics of these different studies.
The fragment-based MP2 approach used by Liu et al. reports $1.80 \times 10^{-9}$ \si{\meter\squared\per\second} for classical MD and $2.27 \times 10^{-9}$ \si{\meter\squared\per\second} for PIMD~\cite{Liu2022/10.1021/acs.jpca.2c00601}.
Here, the PES itself is different due to a difference in the basis set and the use of fragmentation, so a strict match should not be expected.
Further investigation would be required to determine whether this improved performance is indeed due to the denser basis set or due to some cancellation of errors stemming from the MP2 method and the fragment-based approach.

\begin{figure}
\centering
\includegraphics{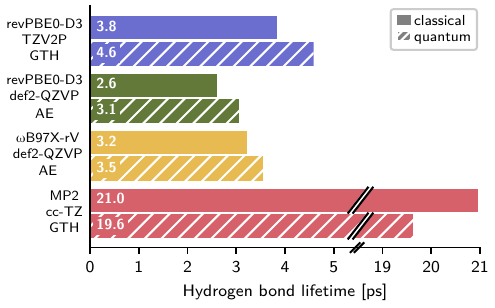}
\caption{\label{fig:hb-lifetimes}
    Hydrogen bond lifetimes for the 4 different C-NNPs.
    To allow for an easier analysis of the shorter bars, we made a cut in the x-axis between 6 and \SI{18}{ps}, indicated by the double-dashes in the axis and the affected bars.
}
\end{figure}

We further investigate the origins of these differences in the dynamics by looking at the characteristic lifetimes of hydrogen bonds in each trajectory.
The results of these calculations are shown in Figure~\ref{fig:hb-lifetimes}.
There is a strong anti-correlation between diffusion coefficients and hydrogen bond lifetimes, as could be expected.
Strong hydrogen bonds have longer lifetimes, which corresponds to slower diffusion.
Equivalently, weak diffusion means that a broken hydrogen bond has a higher chance to recombine again, as the two molecules remain close to each other~\cite{Luzar1996/10.1038/379055a0}.
Compared to the established \revPBElone{} setup, the larger basis set and the switch to an all-electron potential (further details in Figure~\ref{si-fig:hb_lifetimes}) in \revPBEQ{} result in shorter-lived hydrogen bonds.
Similar to the results for the diffusion, the consequences of the basis set change are stronger than for the potential.
With the same setup, the \Bninetylone{} functional yields roughly 20\% higher lifetimes.
For all three DFT setups, including NQEs increases the lifetimes, with \Bninetylone{} showing the smallest effect.
For \MPtwo{}, we find very long lifetimes of around \SI{20}{ps} with a quantum effect that goes in the opposite direction, decreasing the time.

Through the lens of competing NQEs on hydrogen bonding~\cite{Habershon2009/10.1063/1.3167790}, we can see how the changes in the PESs due to more tightly converged settings or differences between functionals manifest in the resulting properties of water.
The geometry of the hydrogen bonds shows both their strengthening due to the intramolecular NQE and their weakening due the intermolecular NQE, with the compensation being similar across the DFT methods.
For MP2, the intramolecular NQE is more pronounced.
The net NQE is then reflected in diffusion and hydrogen bond kinetics.
For \revPBElone{}, there is modest slowdown, with the tighter settings yielding a somewhat smaller effect.
For \Bninety{}, the effect is the smallest, indicating near-complete compensation.
MP2 shows the opposite NQE, with both diffusion and hydrogen bond decay speeding up.
This result should be interpreted with caution, though, due to how overstructured and slow MP2 water is, at least with these settings.

\subsection{Discussion}

\begin{figure}
\centering
\includegraphics{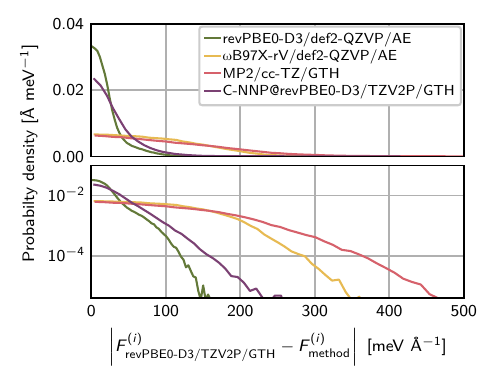}
\caption{
    The distributions of absolute differences in the force components $F^{(i)}$ between \revPBET{} and the three other DFT and post-HF methods, as well as the C-NNP trained to fit \revPBET{}.
    The top panel shows the data on a linear scale, while the bottom panel uses a logarithmic scale on the y-axis to highlight the tails of the distributions.
    The comparison was done on a test set comprising 300 64-molecule bulk liquid water configurations.
}
\label{fig:force_comparison}
\end{figure}

The changes in static and dynamic physical observables discussed above are expressions of changes in the PES of the underlying method.
However, as Figure~\ref{fig:force_comparison} illustrates, it can be difficult to quantify how differences between various PESs affect physical observables.
There, we plot the distribution of absolute differences of force components between \revPBET{} and the other reference methods on a dataset comprising 300 64-molecule structures of bulk liquid water from $NVT$ and $NpT$ trajectories.
As could perhaps be expected, the deviations of \revPBEQ{}, the same functional with tighter settings, are the smallest.
On the other hand, \Bninety{} has considerably higher deviations, despite giving quite similar results to \revPBEQ{}.
The distribution for \MPtwo{} is only slightly wider, even though it produces physical observables drastically different from those of the DFT methods.
Notably, comparing these reference calculations with MLP predictions, the distribution of deviations for C-NNP@\revPBET{} is similar in shape and magnitude (marginally wider, in fact) to that for \revPBEQ{}.
This is despite the fact that in tests comparing actual performance in structure generation using MD, the C-NNP is difficult to distinguish from its reference method~\cite{Schran2020/10.1063/5.0016004}.
This further questions the significance of reporting model errors.
The cause of the differences between model predictions and reference calculations is the epistemic uncertainty of the MLP, which does not affect statistical physical observables in the same way as systematic differences between electronic structure methods, even if the deviations are of a comparable magnitude.

\section{Conclusions}

In this work, we trained four C-NNPs to different reference electronic structure methods using an established training dataset of aqueous systems that we extended to improve the density prediction and enable constant pressure simulations.
After verifying that the transfer of the QbC-generated training dataset to different reference methods does not degrade the models' performance, we used the MLPs to calculate static and dynamic properties of water.
The main focus was to compare two setups for \revPBElone{}, one that used well-established settings with a triple-$\mathrm{\zeta}$ basis set and pseudopotentials, and one that used a quadruple-$\mathrm{\zeta}$ basis set with an all-electron potential and tighter SCF convergence settings.
The other two setups were the range-separated hybrid functional \Bninetylone{} with the same highly converged settings, and MP2 using a moderately-sized triple-$\mathrm{\zeta}$ basis set and pseudopotentials.

When evaluating the models on an independent test set, we found that inaccuracies stemming from insufficient convergence in electronic structure calculations can lead to an increase in test set errors and we showed how using the GAPW method instead of GPW reduces these issues considerably.
In comparisons of the physical observables calculated from MD trajectories, we consistently obtained considerably different results for \revPBET{} and \revPBEQ{} and found that the switch to a larger basis set was more consequential than the switch to an all-electron potential.
With the exception of the pressure--density curve, bringing the basis set closer to the convergence limit meant moving further away from experimental findings.
With a quadruple-$\mathrm{\zeta}$ basis set and all-electron potentials, \Bninetylone{} emerged as the superior functional, although it was not able to match the experiments as closely as \revPBET{}.
Moreover, we confirmed that the use of \MPtwo{} with a triple-$\mathrm{\zeta}$ basis set results in a severe overestimation of the hydrogen bonding strength, leading to highly structured water with insufficient diffusion.

Our results challenge the prevailing notion that triple-$\mathrm{\zeta}$ basis sets and pseudopotentials are sufficient to accurately reveal the true performance of a given exchange--correlation functional for static and dynamic properties of aqueous systems.
Instead, our results suggest that the impressive agreement of \revPBET{} with the experimental data is not due to \revPBElone{} being inherently excellent for water simulations, but rather due to some fortuitous cancellation of the errors of the functional itself and its numerical realization.
This gives an additional perspective on previously identified cancellations within the \mbox{revPBE0} functional viewed through the lens of the many-body expansion~\cite{Riera2019/10.1039/C9SC03291F}.
Importantly, we offer a practical solution in the form of MLPs trained to highly converged electronic structure calculations that only need a modest number of expensive single-point calculations.
This is enabled by uncertainty quantification using committee models, active learning, and the transferable nature of the resulting training sets.
This approach allows one to easily consider multiple methods with converged settings and select the one that offers good performance for the studied system.

In comparison with our MP2 results, the fragment-based approaches by Willow et al.~\cite{Willow2015/10.1038/srep14358} and Liu et al.~\cite{Liu2022/10.1021/acs.jpca.2c00601} show a substantially better match to experiment.
This difference must be due to the differences in basis set, pseudopotential, or fragmentation, though the specifics were not investigated here or elsewhere.
It is not clear from these results what the performance of fully converged MP2 would be for liquid water.
The transfer learning strategy used by Chen et al.~\cite{Chen2023/10.1021/acs.jctc.2c01203}, which required only a moderate number of energies of 12-molecule periodic water structures to train an MLP at the level of correlated electronic structure methods (AFQMC and CCSD(T)) yields a good match to experimental data.
Consistent with the MB-pol model~\cite{Medders2014/10.1021/ct5004115, Zhu2023/10.1021/acs.jctc.3c00326}, their results indicate that an accurate description of aqueous systems can be obtained based on the CCSD(T) level of theory.
One of the cited approaches, our work, or some combination thereof could offer a pathway to accurate models based on correlated electronic structure close to the basis set limit for liquid water and more complex condensed-phase aqueous systems.

Overall, our results highlight both the importance of precise ab initio calculations and the role MLPs can play in enabling fully converged MD simulations based on popular electronic structure methods.

\section*{Associated content}

\subsection*{Data availability statement}

The input scripts, training datasets and models underlying this study are openly available in Zenodo at \url{http://doi.org/10.5281/zenodo.20718510}.

\subsection*{Supporting information}

Additional details on construction of training dataset and validation of C-NNPs.
Insights into the role of the pseudopotential.
Additional RDFs.
Further discussion about convergence settings and their role in model training.

\begin{acknowledgments}

The authors thank Ngoc Lan Le Nguyen, Tomáš Martinek, Pavol Šimko and Christoph Schran for their valuable feedback on the manuscript.
The authors acknowledge support from the Czech Science Foundation, project No. 21-27987S.
This work was supported by the Ministry of Education, Youth and Sports of the Czech Republic through the e-INFRA CZ (ID: OPEN-24-70 and OPEN-28-19).

\end{acknowledgments}

\putbib

\end{bibunit}


\clearpage

\setcounter{section}{0}
\setcounter{equation}{0}
\setcounter{figure}{0}
\setcounter{table}{0}
\setcounter{page}{1}

\renewcommand{\thesection}{S\arabic{section}}
\renewcommand{\theequation}{S\arabic{equation}}
\renewcommand{\thefigure}{S\arabic{figure}}
\renewcommand{\thepage}{S\arabic{page}}
\renewcommand{\citenumfont}[1]{S#1}
\renewcommand{\bibnumfmt}[1]{$^{\rm{S#1}}$}
\newcommand{\NLL}{\mathrm{NLL}}

\title{Supporting information for: \mytitle}
{\maketitle}

\onecolumngrid
\fontsize{12}{14}\selectfont

\begin{bibunit}



\section{Adapting the Training Dataset}
\label{si-sec:training_data}

\nocite{Schran2020/10.1063/5.0016004} 
\begin{figure}
\centering
\includegraphics[width=\columnwidth]{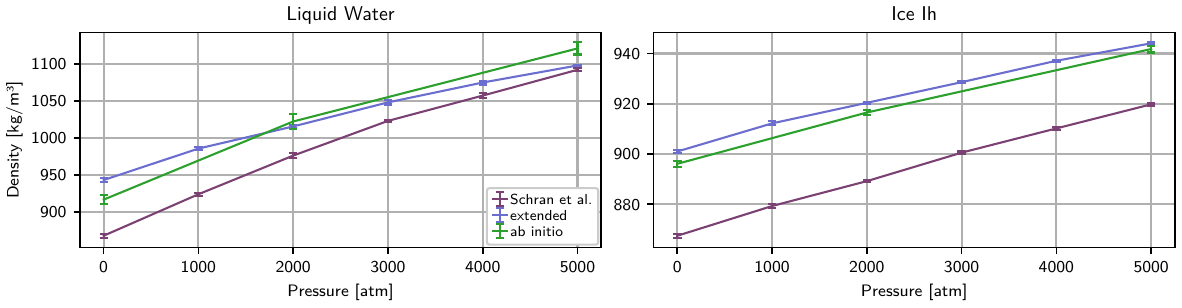}
\caption{\label{si-fig:density-generations}
    Comparison of pressure--density curves from ab initio simulations and C-NNPs trained on two different training sets.
    The setup for the ab initio simulations and the two training sets is \revPBET{}.
    Schran et al. corresponds to the dataset originally published in reference \citenum{Schran2020/10.1063/5.0016004}, ``extended'' to the version that we extended by 150 $NpT$ structures.
    The left panel displays the pressure--density curve for bulk liquid water, the right panel for ice \ih.
    Error bars were obtained using block averaging.
}
\end{figure}

The water training dataset published by Schran et al.~\cite{Schran2020/10.1063/5.0016004} consists of 814 structures of bulk liquid water, ice, and a slab with a liquid--vacuum interface sampled from the canonical ensemble ($NVT$).
Committee neural network potentials (C-NNPs) trained on this dataset are able to reproduce many physical observables from ab initio simulations with the same reference method.
However, as Figure~\ref{si-fig:density-generations} shows, the models underestimate the density of both liquid water and ice \ih.
Therefore, we extended the training dataset by adding structures from the isobaric--isothermal ensemble ($NpT$) simulated with the C-NNP at pressures of 1, 2000 and 5000 \si{atm} using a query by committee (QbC) workflow~\cite{Schran2020/10.1063/5.0016004}.
In total, we added 110 64-molecule structures of liquid water and 40 92-molecule structures of ice \ih.
Although these configurations were obtained from simulations using a model that produces the wrong density, including them into the training set with the correct energies and forces from ab initio calculations considerably improves the density predictions.
For liquid water (shown in the left panel of Figure~\ref{si-fig:density-generations}), the density does not match the results from ab initio simulations perfectly, but this is likely due to insufficient convergence of the expensive ab initio trajectories.

\section{Resource requirements}
\label{si-sec:resources}

\begin{table}
    \centering
    \begin{tabular}{l|rrr}
         & wall time & memory & resources \\
         \hline
        \revPBET & 0:04 & 18\,GB & 1 EPYC 7301 node\\
         \revPBEQ & 0:14 & 35\,GB & 1 EPYC 7301 nodes\\
         \Bninety & 0:34 & 123\,GB & 1 EPYC 7301 node\\
         \MPtwo & 0:28 & 3.2\,TB & 16 EPYC 7H12 nodes\\
    \end{tabular}
    \caption{
        \label{si-tab:liquid}
        Computational costs for a single point calculation of a 64-molecule structure of bulk liquid water.
    }

    \centering
    \begin{tabular}{l|rrr}
         & wall time & memory & resources \\
         \hline
        \revPBET & 0:05 & 24\,GB & 1 EPYC 7301 node\\
         \revPBEQ & 0:20 & 55\,GB & 1 EPYC 7301 node\\
         \Bninety & 0:54 & 185\,GB & 1 EPYC 7301 node\\
         \MPtwo & 4:20 & 3.1\,TB & 16 EPYC 7H12 nodes\\
    \end{tabular}
    \caption{
        \label{si-tab:ice}
        Computational costs for a single point calculation of a 96-molecule structure of ice \ih.
    }

    \centering
    \begin{tabular}{l|rrr}
         & wall time & memory & resources \\
         \hline
        \revPBET & 0:11 & 54\,GB & 1 EPYC 7301 node\\
         \revPBEQ & 0:28 & 760\,GB & 2 EPYC 7H12 nodes\\
         \Bninety & 0:35 & 952\,GB & 4 EPYC 7H12 nodes\\
    \end{tabular}
    \caption{
        \label{si-tab:slab}
        Computational costs for a single point calculation of a 216-molecule slab of liquid water.
        Note the switch to larger nodes compared to liquid water and ice calculations for the highly converged setups.
        One EPYC 7031 node contains two AMD EPYC 7031 16-core CPUs.
        One EPYC 7H12 node contains two AMD EPYC 7H12 64-core CPUs.
        The wall time is given in the format hours:minutes.
    }
\end{table}

Tables \ref{si-tab:liquid}, \ref{si-tab:ice}, and \ref{si-tab:slab} show the computational costs for a single point calculation for the different methods for one configuration of liquid water, ice \ih{}, and slab, respectively.
In each calculation, the SCF was initialized with converged Kohn--Sham orbitals from corresponding GGA calculations to speed up convergence.
Please note that an important reason for the increased demands in CPU time and memory of \Bninety{} compared to \revPBEQ{} is that \Bninety{} requires a lower Schwarz inequality threshold, which controls the screening of near-field electronic repulsion integrals.
The lower threshold results in a substantially higher number in integrals during the SCF procedure.
Furthermore, during the calculation of the correlation energy for MP2, an MPI group size of one process was used for liquid water, and two processes for ice \ih{}.
This change results in a lower memory requirement but longer wall time for ice \ih{} calculations.

\section{Complete Test Sets}
\label{si-sec:testsets}

\begin{figure}
\centering
\includegraphics[width=.49\columnwidth]{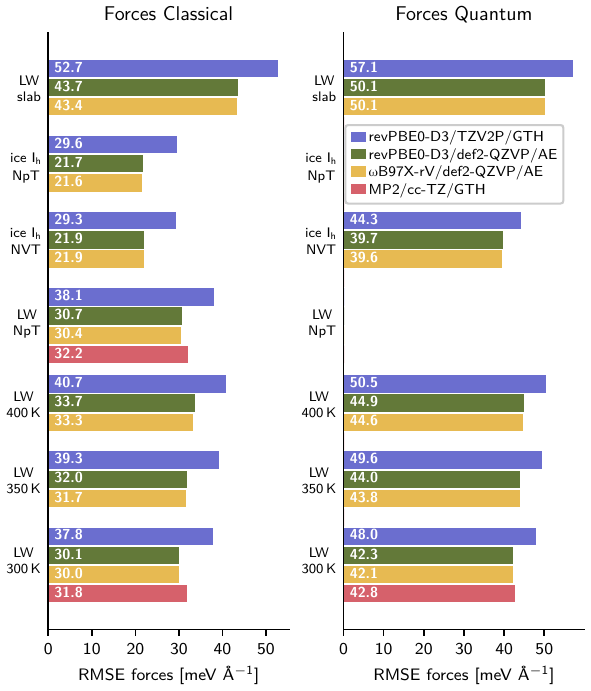}
\includegraphics[width=.49\columnwidth]{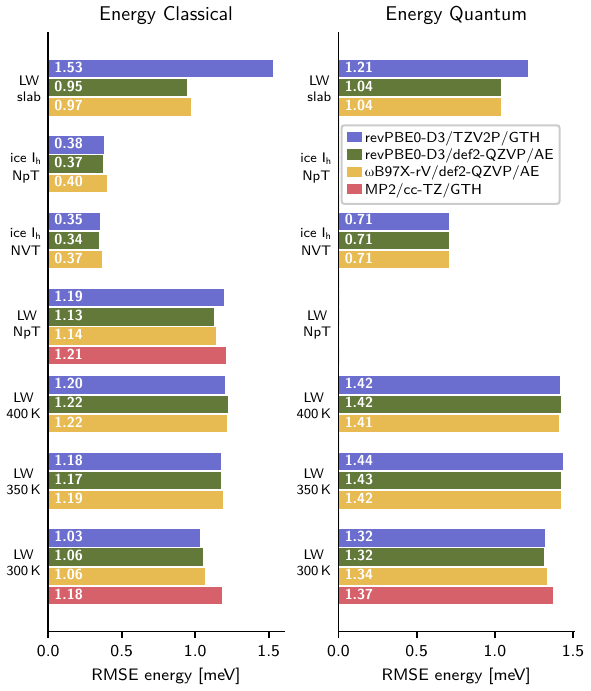}
\caption{\label{si-fig:testsets}
    Test set errors of the four main models.
    The two panels on the left display the force RMSEs for classical and quantum structures and complete Figure~\ref{fig:testset} of the main paper.
    The two panels on the right display the energy RMSEs for the same test sets.
}
\end{figure}

Figure~\ref{si-fig:testsets} shows the test set errors of the four main models for both energy and forces.
The two panels on the left display the force errors and complete Figure~\ref{fig:testset} of the main article.
The additional test sets are fully consistent with the observations made in the main paper.
The energy RMSEs displayed in the right panels show that for most test sets the energy errors are very consistent.
The increase in test errors for the \revPBET{} datasets due to the egg box effect, which we observed in the force errors, is absent in most test sets.
The only exception is the slab test sets, where \revPBET{} has a considerably higher error.
The reason why the slab structures are an exception and why the energy error is higher for classical slab configurations than for quantum slab configurations remains unclear.

\section{Role of Basis Sets and Pseudopotentials}
\label{si-sec:pseudopotential}

In the main paper, we compared a standard setup for \revPBElone{} with one that was highly converged.
The main differences between the two setups were the more extensive basis set and the use of an all-electron potential instead of GTH pseudopotentials~\cite{Goedecker1996/10.1103/PhysRevB.54.1703} for the highly converged setup.
More details can be found in the computational details (Section~\ref{sec:comp-details}) of the main article.
In this section, we will show in greater detail the individual contributions of the basis set and potential by presenting the results for a model trained to fit the \revPBEQGTH{} setup and comparing them with the results of the \revPBET{} and \revPBEQ{} models already shown in the main paper.

\subsection{Radial Density Function}

\begin{figure}
\centering
\includegraphics[width=\columnwidth]{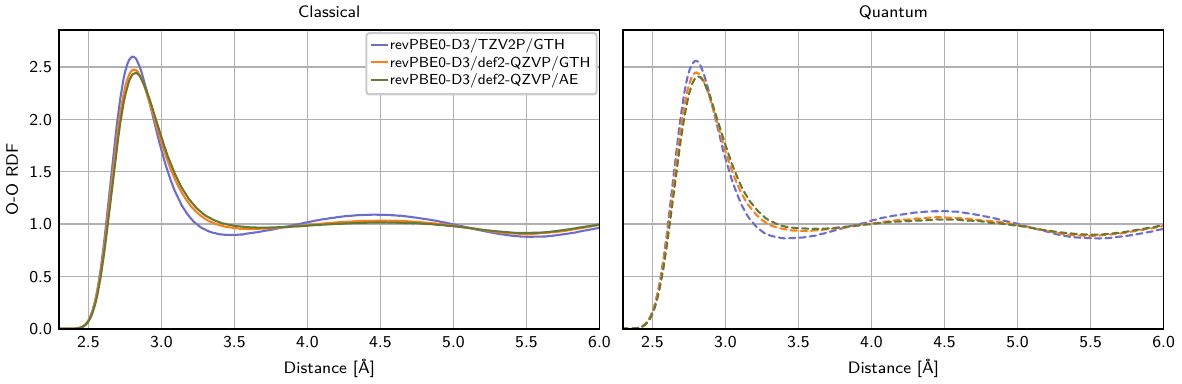}
\caption{\label{si-fig:rdf_oo}
    The oxygen-oxygen RDF for the 3 \revPBElone{} C-NNPs.
    The results from the classical MD simulations are shown in the left panel, the PIMD results are shown in the right panel.
}
\end{figure}

Figure~\ref{si-fig:rdf_oo} shows the oxygen-oxygen radial density function (RDF) for classical and quantum simulations.
The RDFs of the \revPBEQGTH{} model are very similar to those of the \revPBEQ{} model, while there are considerable differences from the \revPBET{} RDFs.
This indicates that for distances around the first coordination shell and beyond, the pseudopotentials are a good approximation to the physically accurate all-electron potentials.
Considering that oxygen and especially hydrogen are quite small atoms, and only a few core electrons of the oxygen atoms are included in the pseudopotential, this strong match can be expected.
On the other hand, changing the basis set has a considerable effect on the RDF, which is discussed in greater detail in the main article.

\subsection{Covalent and Hydrogen Bonds}

\begin{figure}
\centering
\includegraphics[width=.5\columnwidth]{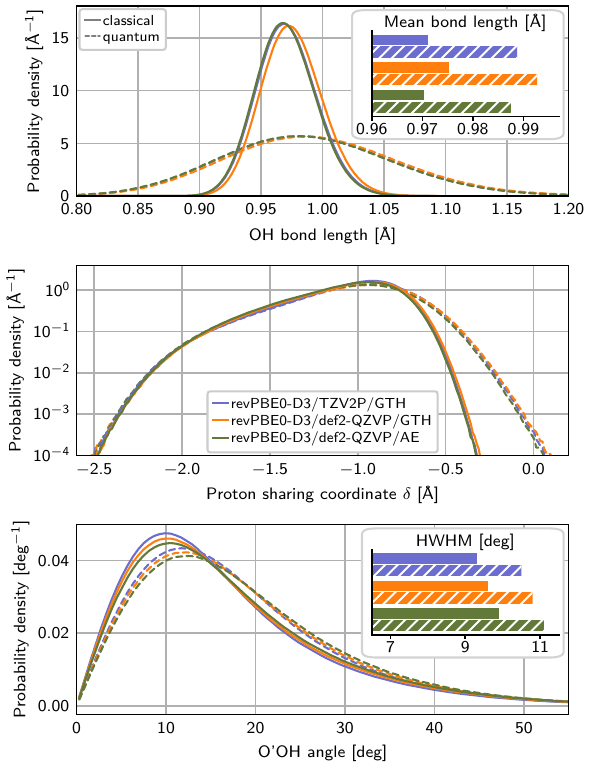}
\caption{\label{si-fig:bonds_angles}
    Analysis of the structure of covalent and hydrogen bonds.
    Top panel: The distribution and means of lengths of covalent oxygen-hydrogen bonds for classical (solid lines/bars) and PI (dashed lines/bars) MD run with the 3 different C-NNPs.
    The main plot shows the distributions of bond lengths and the inset the mean bond lengths.
    Middle panel: The distributions of the proton sharing coordinate.
    Bottom panel: The distributions of the hydrogen bond angle.
    The inset compares the half width at half maximum of the different methods.
}
\end{figure}

Figure~\ref{si-fig:bonds_angles} shows an analysis of the covalent and hydrogen bonds of the three \revPBElone{} models, analogous to Figure~\ref{fig:bonds_angles} in the main article.
The distribution of bond lengths shown in the top panel shows only miniscule differences between \revPBET{} and \revPBEQ{}.
While it can be expected that \revPBEQGTH{}, which is a mix of these two setups, gives a similar distribution, we see a clear shift towards longer covalent bonds, especially for the classical system.
We conclude that in our findings the improved basis set increases the bond lengths, whereas the all-electron potential has the opposite effect, leading to a high similarity between the state-of-the-art and the fully converged setup.
This indicates that in close proximity to the nuclei, there is a measurable but not disastrous difference between pseudopotentials and all-electron potentials.
In the hydrogen bonds, we see less consequential differences between the models.
The $\delta$-coordinate plotted in the centre panel of Figure~\ref{si-fig:bonds_angles} displays almost no difference between the three models.
The distribution of the hydrogen bond angles in the bottom panel shows that the distribution of \revPBEQGTH{} is in the middle of the other two models, illustrating that both the increased basis set and the upgraded potential impact the hydrogen bonding.

\subsection{Pressure--Density Curve}

\begin{figure}
\centering
\includegraphics[width=\columnwidth]{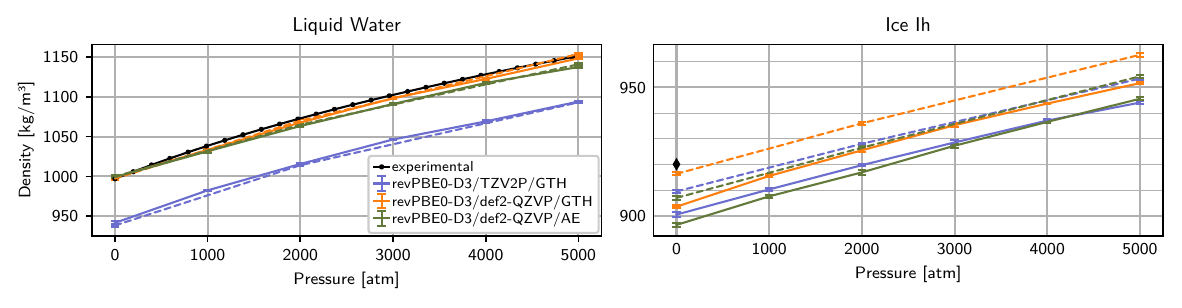}
\caption{\label{si-fig:density}
    The pressure--density curve of liquid water (left) and ice \ih (right) for the three different C-NNPs.
    The black curve in the left panel and the black diamond in the right panel shows the experimental reference~\cite{Grindley1971/10.1063/1.1675455, Sanz2004/10.1103/PhysRevLett.92.255701} for each system.
    The error bars indicate statistical errors obtained by block averaging.
    As in the other plots, solid lines are used for classical MD and dashed lines for PIMD.
}
\end{figure}

Figure~\ref{si-fig:density} shows the pressure--density curve for the three \revPBElone{} models.
For liquid water, which is shown in the left panel, \revPBEQGTH{} and \revPBEQ{} are in almost perfect agreement with each other and the experimental reference curve~\cite{Grindley1971/10.1063/1.1675455}.
Their density is consistently around \SI{50}{\kilo\gram\per\meter\cubed} higher than the \revPBET{} density.
The differences between classical and quantum simulations are small for all models.
In the right panel of Figure~\ref{si-fig:density} the pressure--density curves for ice \ih are plotted and display a different behaviour than the bulk liquid water curves.
Here, the density corresponding to the \revPBEQGTH{} models is slightly higher than the other two \revPBElone{} models, with discrepancies of less than \SI{10}{\kilo\gram\per\meter\cubed}.

\subsection{Diffusion Coefficient}

\begin{figure}
\centering
\includegraphics[width=.5\columnwidth]{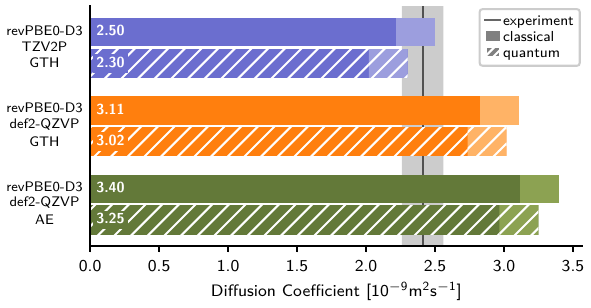}
\caption{\label{si-fig:diffusion}
    Comparison of the diffusion coefficients for the different C-NNPs in classical and quantum MD.
    The pale colours at the end of each bar signals the magnitude of the finite size corrections added to the calculated values.
    The gray vertical area indicates the experimental reference value~\cite{Holz2000/10.1039/b005319h}.
}
\end{figure}

Figure~\ref{si-fig:diffusion} shows the diffusion coefficients for the three \revPBElone{} models, analogous to Figure~\ref{fig:diffusion} in the main paper.
The diffusion coefficient for \revPBEQGTH{} lies in between the two other \revPBElone{} models, with smaller discrepancies to the highly converged setup.
This indicates that while both changes have a meaningful impact, the effect of the increased basis set is stronger than the change in potential.
For all three models, we observe a slightly decreased diffusion coefficient when including nuclear quantum effects.

\subsection{Hydrogen Bond Lifetimes}

\begin{figure}
\centering
\includegraphics[width=.5\columnwidth]{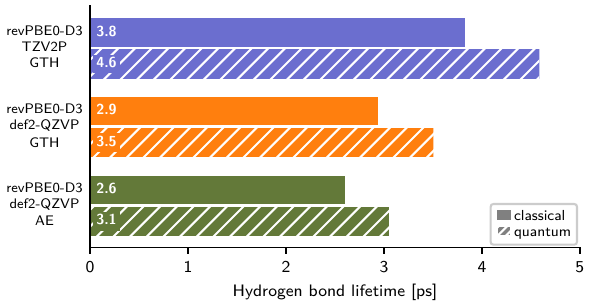}
\caption{\label{si-fig:hb_lifetimes}
    Hydrogen bond lifetimes for the 3 different \revPBElone{} C-NNPs.
}
\end{figure}

Figure~\ref{si-fig:hb_lifetimes} shows the hydrogen bond lifetimes for the three \revPBElone{} models, analogous to Figure~\ref{fig:hb-lifetimes} in the main paper.
As with the diffusion constant, the results of \revPBEQGTH{} lie in between the results of the other two models, with a stronger similarity to the \revPBEQ{} setup.
This further highlights the strong connection between hydrogen bond lifetimes and diffusion.

\subsection{Conclusion}

We conclude that the results of \revPBEQGTH{} lie between those of \revPBET{} and \revPBEQ{} for most of the physical observables that we investigated.
This indicates that the increased size of the basis set and the all-electron potential impact the water systems in similar ways.
However, for observables where long-range interactions play a considerable role, the effects of increasing the size of the basis set were stronger than those of switching from a pseudopotential to an all-electron potential.
This also highlights the high quality of the GTH pseudopotential and their ability to reproduce the behaviour of the physically accurate all-electron potentials.
The only exception was the distribution of covalent bonds, where the two changes cancel each other out.

\section{Other Radial Density Functions}
\label{si-sec:rdfs}

\begin{figure}
\centering
\includegraphics[width=\columnwidth]{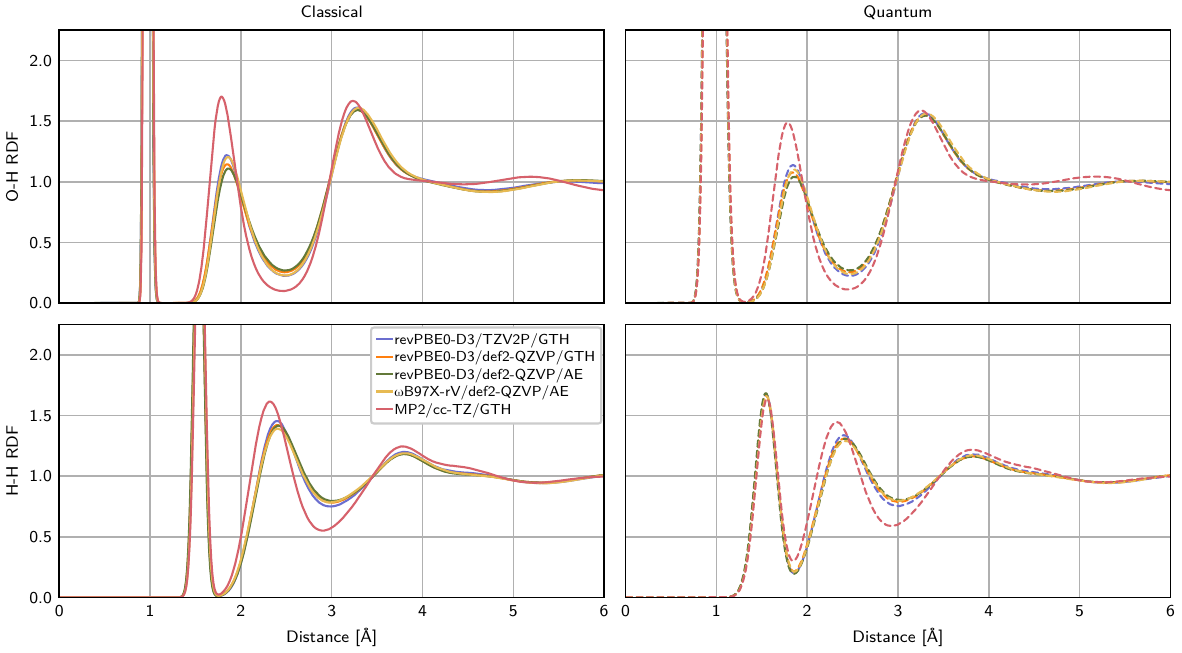}
\caption{\label{si-fig:rdf_others}
    The oxygen-hydrogen (top panel) and hydrogen-hydrogen (bottom) RDF for the five C-NNPs discussed in the main article and here.
    The results from the classical MD simulations are shown in the left panels, the PIMD results are shown in the right panels.
}
\end{figure}

Figure \ref{si-fig:rdf_others} shows the oxygen-hydrogen (top panels) and hydrogen-hydrogen (bottom panels) RDFs for the four models from the main paper, as well as the \revPBEQGTH{} model.
The general trends, which we have already observed for the oxygen-oxygen RDF in Figure \ref{fig:rdf} of the main article, are confirmed here as well.
In particular, the RDFs for the \MPtwo{} model are notably more structured than the RDFs of the other models, with peaks and valleys occurring at a shorter distance.
Even for longer distances, considerable differences in the structure can still be observed.
When comparing \revPBET{} and \revPBEQ{}, the trend of a smoother RDF for the highly converged setup is confirmed, especially for the first inter-molecular peak of the O-H RDF.
However, the trend that the \Bninety{} and \revPBEQ{} models display a similar RDF as observed for the O-O RDF is broken.
Now, the O-H RDF of \Bninety{} shows greater similarity with \revPBET{}.

\section{Under-converged Datasets}
\label{si-sec:egg-box}

In this section, we will further investigate under-converged datasets and their effect on model testing and training.
Under-converged in this case does not refer to the type of Gaussian basis sets or potentials, but to convergence settings, plane wave cutoffs and calculation methods.
We will look into the net forces of systems, the strength of the sum over all forces in the system, which should be 0.
They have been shown to be a useful tool for finding unreliable electronic structure calculations~\cite{Kuryla2025/10.48550/arXiv.2510.19774}.
One of the primary causes for errors in calculations using plane wave basis sets is the so called ``egg box effect''~\cite{Durham2025/10.1088/2516-1075/adc056}.
The egg box effect results in different total energies of identical systems based on their location relative to the real space computational grid of the plane wave basis set.
In case of a low plane wave energy cutoff, this grid is coarse, resulting in inconsistencies.

\subsection{Net Forces}

\begin{figure}
\centering
\includegraphics[width=\columnwidth]{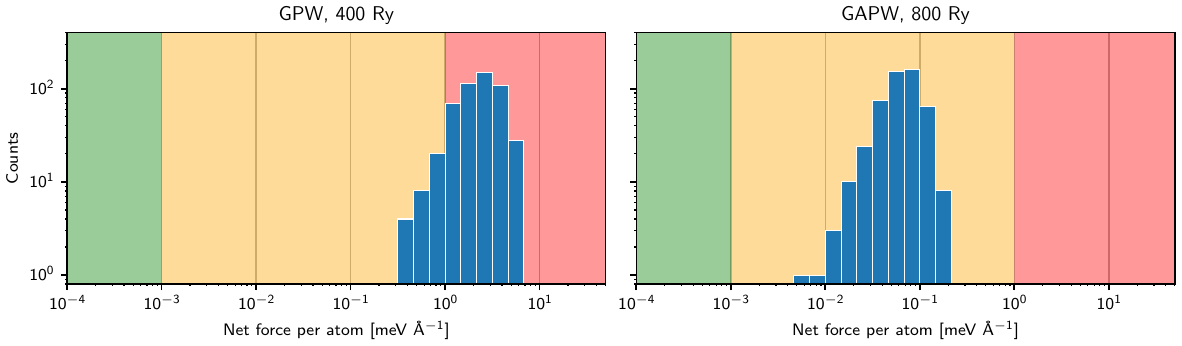}
\caption{\label{si-fig:net_forces_comp}
    The distribution of net forces per structure for the classical liquid water NpT test set.
    In the left panel the standard \revPBET{} setup (with GPW and a \SI{400}{Ry} plane wave cutoff) as used in the main paper is shown, in the right panel instead the GAPW method with a \SI{800}{Ry} cutoff was employed.
    The background colours correspond to the thresholds used in reference~\citenum{Kuryla2025/10.48550/arXiv.2510.19774}.
    The red colour above \SI{1}{\milli\eV\per\angstrom} indicates problematic errors, the orange colour between $10^{-3}$ and $1$ \si{\milli\eV\per\angstrom} corresponds to acceptable errors and the green area below $10^{-3}$ \si{\milli\eV\per\angstrom} means negligible errors.
}
\end{figure}

Figure~\ref{si-fig:net_forces_comp} shows the distribution of net forces per atom for two versions of the classical liquid water NpT test set, which consists of 500 structures and was one of the test sets used for Figure~\ref{fig:testset} of the main paper.
The area with the red-shaded background colour indicates a magnitude of net forces that is indicative of problems in the calculation setup.
The left panel shows the net forces when calculating the test set with the standard \revPBET{} setup, and a large part of the distribution is in the red zone.
The right panel of Figure \ref{si-fig:net_forces_comp} shows the result for the test set with the same basic DFT setup, but instead of the Gaussian and plane waves (GPW) method~\cite{Lippert1997/10.1080/002689797170220}, it used the Gaussian and augmented plane waves (GAPW) method~\cite{Lippert1999/10.1007/s002140050523} in addition to an increased plane wave cutoff.
This change in the setup considerably improves the net forces by decreasing them by more than one order of magnitude.
For the 64-atom structures, this means a reduction in the total net forces per structure from \SI{486}{\milli\eV\per\angstrom} to \SI{13}{\milli\eV\per\angstrom}.
This is lower than the \SI{40}{\milli\eV\per\angstrom} that Kuryla et al.~\cite{Kuryla2025/10.48550/arXiv.2510.19774} achieved for a similar water setup in CP2K with a plane wave cutoff of \SI{1200}{Ry}.
It shows that the GAPW method is an important tool in improving the quality of datasets calculated in CP2K.
This improvement in the quality of the test data set leads to an immediate improvement in the test score.
When testing the \revPBET{} model on the original dataset, we obtain a force RMSE of \SI{38.1}{\milli\eV\per\angstrom}.
Using the same model on the improved test set, we get a force RMSE of \SI{33.8}{\milli\eV\per\angstrom}, despite technically using a different reference method from the training data.

\begin{figure}
\centering
\includegraphics[width=\columnwidth]{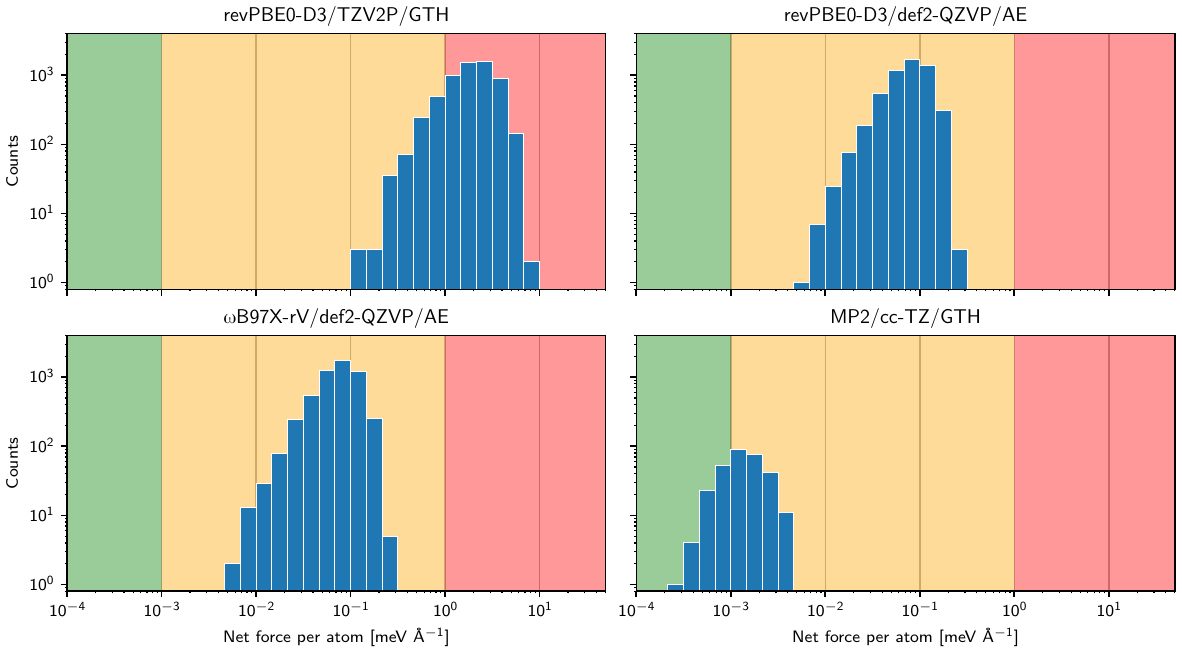}
\caption{\label{si-fig:net_forces_full_trainset}
    The distribution of net forces per structure for the full test sets for the four main reference methods.
    The background colours correspond to the thresholds used in reference~\citenum{Kuryla2025/10.48550/arXiv.2510.19774}.
    The red colour above \SI{1}{\milli\eV\per\angstrom} indicates problematic errors, the orange colour between $10^{-3}$ and $1$ \si{\milli\eV\per\angstrom} corresponds to acceptable errors and the green area below $10^{-3}$ \si{\milli\eV\per\angstrom} means negligible errors.
}
\end{figure}

Figure~\ref{si-fig:net_forces_full_trainset} shows the distribution of net forces per atom for the collective test sets for all four electronic structure calculation setups.
The dataset is the full test set established by Schran et al.~\cite{Schran2020/10.1063/5.0016004} extended by a set of 500 structures from NpT simulations of bulk water, which was already used in Figure~\ref{si-fig:net_forces_comp}.
The two top panels display the distribution of the two \revPBElone{} datasets, further proving that the GAPW method reduces inconsistencies in DFT calculations.
The distributions of \revPBEQ{} (top right) and \Bninety{} (bottom left) are very similar, highlighting that the improvement is largely independent from the exchange--correlation functional.
The distribution of the \MPtwo{} setup (bottom right) is shifted further to the left by almost two orders of magnitude, with almost half of the structures scoring below the $10^{-3}$ \si{\milli\eV\per\angstrom} threshold.
The MP2 distribution is smaller because only a subset of the full test set could be calculated due to the expensive nature of bulk MP2 calculations.

\subsection{Effect of the Egg Box Effect on Training}

\begin{figure}
\centering
\includegraphics[width=.5\columnwidth]{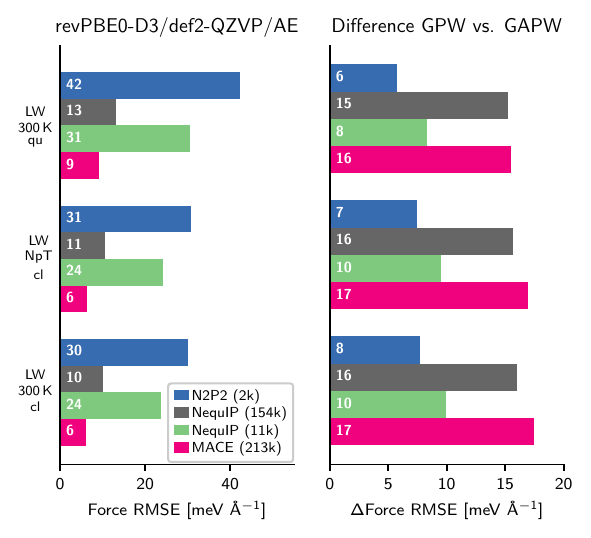}
\caption{\label{si-fig:egg_box}
    The left panel shows the force RMSE of four different models architectures trained and tested on the \revPBEQ{} datasets.
    The right panel shows the difference in force RMSEs between \revPBET{} models and \revPBEQ{} models when tested on their respective test sets.
    In the legend, the number in brackets is the number of trainable parameters in each model.
}
\end{figure}

We have highlighted how the egg box effect is a primary source of error in calculations using plane wave basis set and how these errors can be diminished by the GAPW method.
Furthermore, we showed how these errors lead to an overestimation in a model's test error as the evaluation of a single structure is more strongly affected by these problems than the training of a model with a large set of structures.
Here, we investigate further how this error affects the training by comparing models with different architectures and capacities on the same training datasets.
In addition to the N2P2~\cite{Singraber2019/10.1021/acs.jctc.8b00770, Singraber2019/10.1021/acs.jctc.8b01092} models used in the main article, NequIP~\cite{Batzner2022/10.1038/s41467-022-29939-5} and MACE~\cite{Batatia2022/MACE} models were trained using the \revPBET{} and \revPBEQ{} training sets.
For NequIP, we trained one model with commonly used settings, which amounts to a total of 154\,000 trainable parameters and one extremely minimalistic model containing just 11\,000 parameters.
The MACE model uses some common settings, resulting in 213\,000 parameters.
The left panel of Figure~\ref{si-fig:egg_box} shows the errors of the four models on different liquid water test sets calculated with \revPBEQ{}.
As expected, the modern message-passing models MACE and NequIP can reduce the force error considerably compared to N2P2 and the minimalistic NequIP model performs worse than the baseline NequIP model.
The right panel of Figure~\ref{si-fig:egg_box} shows the difference in test errors between the errors from \revPBET{} models and \revPBEQ{} models tested on their respective test sets.
As we have seen in Figure~\ref{fig:testset} of the main paper, the performance of the model is largely independent from the exact potential energy surface.
Instead, the presence of errors in the datasets due to the egg box effect is determining the difference in test set errors of the \revPBET{} and \revPBEQ{} models.
It is clearly evident that this difference is larger for models, which have a high capacity and perform very well on the test set, than for models, which perform worse.
We believe that the origin of this difference lies in the ability of a model to fit the noise in the training data stemming from the egg box effect.
Low-capacity models such as N2P2 or the minimalistic NequIP setup smooths out the noise in the \revPBET{} dataset, whereas high-capacity models are able to fit the training data so well that they fit some of the noise of the egg box effect.

\FloatBarrier
\section*{References}

\putbib

\end{bibunit}

\end{document}